\documentclass[10pt,journal,compsoc]{IEEEtran}
\usepackage{cite}
\usepackage{amsmath,graphicx}
\usepackage{epstopdf}
\usepackage{ragged2e}
\usepackage{xcolor}
\usepackage{amssymb}
\usepackage{pifont}
\usepackage{array}
\usepackage{booktabs}
\usepackage{bm}
\usepackage{multirow}
\usepackage{graphicx}
\usepackage{color}
\usepackage{float}
\usepackage{stfloats}
\usepackage[misc]{ifsym}
\usepackage{stmaryrd}
\usepackage{grffile}
\usepackage{subfig}
\usepackage{makecell}
\usepackage{threeparttable}
\usepackage{enumitem}
\usepackage{enumerate}
\usepackage{hyperref}
\hypersetup{urlcolor=magenta, citecolor=blue, linkcolor=red, colorlinks=true}
\usepackage{colortbl}
\definecolor{shadow}{cmyk}{.08,0,0,0}
\usepackage{tikz}
\newcommand*{\circled}[1]{\lower.6ex\hbox{\tikz\draw (0pt, 0pt)%
    circle (.4em) node {\makebox[1em][c]{\tiny #1}};}}
\begin{document}

\title{Disentangling Light Fields for Super-Resolution and Disparity Estimation}

\author{Yingqian Wang, Longguang Wang, Gaochang Wu, Jungang Yang, Wei An,\\ Jingyi Yu, \IEEEmembership{Fellow, IEEE}, and Yulan Guo, \IEEEmembership{Senior Member, IEEE}

\IEEEcompsocitemizethanks{
\IEEEcompsocthanksitem
Y. Wang,  L. Wang, J. Yang, W. An, and Y. Guo are with the College of Electronic Science and Technology, National University of Defense Technology, Changsha, P. R. China.
\IEEEcompsocthanksitem
G. Wu is with the State Key Laboratory of Synthetical Automation for Process Industries, Northeastern University, P. R. China.
\IEEEcompsocthanksitem
J. Yu is with the School of Information Science and Technology, ShanghaiTech University, P. R. China.
}
\thanks{This work was partially supported in part by the National Natural Science Foundation of China (Nos. U20A20185, 61972435, 61401474, 61921001). Corresponding author: Jungang Yang. Emails: yangjungang@nudt.edu.cn.}
}

\markboth{}%
{Shell \MakeLowercase{\textit{et al.}}: Bare Advanced Demo of IEEEtran.cls for IEEE Computer Society Journals}

\IEEEtitleabstractindextext{
\begin{abstract}
\justifying
Light field (LF) cameras record both intensity and directions of light rays, and encode 3D scenes into 4D LF images. Recently, many convolutional neural networks (CNNs) have been proposed for various LF image processing tasks. However, it is challenging for CNNs to effectively process LF images since the spatial and angular information are highly inter-twined with varying disparities. In this paper, we propose a generic mechanism to disentangle these coupled information for LF image processing. Specifically, we first design a class of domain-specific convolutions to disentangle LFs from different dimensions, and then leverage these disentangled features by designing task-specific modules. Our disentangling mechanism can well incorporate the LF structure prior and effectively handle 4D LF data. Based on the proposed mechanism, we develop three networks (i.e., DistgSSR, DistgASR and DistgDisp) for spatial super-resolution, angular super-resolution and disparity estimation. Experimental results show that our networks achieve state-of-the-art performance on all these three tasks, which demonstrates the effectiveness, efficiency, and generality of our disentangling mechanism. Project page: \url{https://yingqianwang.github.io/DistgLF/}.
\end{abstract}

\begin{IEEEkeywords}
Light field image processing, feature disentangling, image super-resolution, view synthesis, disparity estimation
\end{IEEEkeywords}}

\maketitle

\IEEEdisplaynontitleabstractindextext

\IEEEpeerreviewmaketitle

\section{Introduction}\label{sec:introduction}

\IEEEPARstart{L}{ight} field (LF) cameras can capture both intensity and directions of light rays, and record 3D geometry in a convenient and efficient manner. By encoding 3D scene cues into 4D LF images (i.e., 2D for spatial dimension and 2D for angular dimension), LF cameras enable many attractive applications such as post-capture refocusing \cite{wang2018selective,bishop2011light}, depth sensing \cite{wang2016depth,park2017robust,PS-RF,CAE}, reflectance estimation \cite{zhou2020shape,wang2017svbrdf,tao2015depth,alperovich2018light}, and foreground de-occlusion \cite{DeOccNet,Mask4D}. With recent development of LF imaging devices (e.g., plenoptic cameras \cite{ng2005light}, coded aperture cameras \cite{sakai2020acquiring}, moving gantries \cite{STFgantry}, and camera arrays \cite{wilburn2005high}), LF image processing has drawn extensive research interests due to its large potential in both academic and industrial communities \cite{wu2017overview}.

Recently, convolutional neural networks (CNNs) have been extensively studied for various LF image processing tasks. However, due to the complex scene structures in real-world scenarios, the spatial and angular information in LF images are highly inter-twined with varying disparities, which introduces great challenges to CNNs to exploit informative cues. To solve this problem, existing methods generally take the redundant LF structure (i.e., by recording similar image contents from different angular views) into consideration, and reduce the dimension of LF images by performing convolutions on neighboring views \cite{LFCNN15,LFCNN17}, epipolar plane images (EPIs) \cite{LFEPICNN17,LFEPICNN18,ShearedEPI,SAAN}, or 3D sub-LFs \cite{EPINET,resLF,MALFRNet,EPI-Shift,AttMLFNet}. Although these methods are effective in handling LF data, their performance is limited due to the under-exploitation of the rich angular information.

Since different views in LF images are regularly arranged in the 2D angular dimension, the performance of CNNs can be enhanced by fully incorporating this structural prior. In this paper, we consider this prior into the deisgn of CNNs and propose a generic disentangling mechanism for LF image processing. Specifically, we organize the input LF into a macro-pixel image (MacPI) and then design a class of convolutions (e.g., spatial$/$angular feature extractors) to achieve domain-specific feature extraction.  After LF disentanglement, we use several specifically-designed modules to fuse the disentangled features for different LF image processing tasks.

Compared to existing LF image processing frameworks, our disentangling mechanism has three remarkable properties. \textbf{First}, our mechanism can well incorporate the LF structure prior and fully use the information from all angular views. \textbf{Second}, since the 4D data can be disentangled into several subspaces by our mechanism, the convolution layers in our networks only need to process features in a single subspace, which makes LF representation learning easier. \textbf{Third}, our mechanism is generic and can be applied to different LF image processing tasks.

This paper is an extension of our previous conference version \cite{LF-InterNet} in which we proposed a spatial-angular interaction network (i.e., LF-InterNet) for LF spatial super-resolution (SR). This work makes the following additional contributions as compared to our preliminary version:

\begin{itemize}
	\item We generalize the spatial-angular interaction mechanism to the disentangling mechanism. The proposed mechanism is generic and more effective in handling LF data than LF-InterNet.
	\item We apply the disentangling mechanism to LF spatial SR by proposing a network named DistgSSR. Our DistgSSR significantly outperforms LF-InterNet with a smaller model size.
	\item We test the generality of the proposed mechanism by developing two networks (namely, DistgASR and DistgDisp) for LF angular SR and disparity estimation. Our Distg-based networks achieve state-of-the-art performance on these two tasks.
\end{itemize}

The rest of this paper is organized as follows. In Section~\ref{sec:RelatedWork}, we briefly review the related work. In Section~\ref{sec:DistgMechanism}, we introduce our disentangling mechanism. The applications of our mechanism in LF spatial SR, LF angular SR, and LF disparity estimation are introduced in Section~\ref{sec:DistgSSR}, \ref{sec:DistgASR}, \ref{sec:DistgDisp}, respectively. Finally, we conclude this paper in Section \ref{sec:Conclusion}.

\section{Related Work}\label{sec:RelatedWork}
LF image processing has been investigated for various tasks in recent decades. Here, we focus on three typical tasks including spatial SR, angular SR, and disparity estimation. Moreover, we briefly review several widely-used CNN architectures for LF image processing, which are closely related to this work.

\subsection{Spatial Super-Resolution}
LF spatial SR, also termed as LF image SR, aims at generating high-resolution (HR) LF images from their low-resolution (LR) inputs. A straight-forward approach to achieve LF spatial SR is to apply single image SR (SISR) methods to each sub-aperture image (SAI) independently. However, directly performing SISR for LF spatial SR cannot produce satisfactory results since the correlation among different views is overlooked. To achieve high performance in LF spatial SR, the information within a single view (i.e., spatial information) and among different views (i.e., angular information) should both be exploited.

Early studies \cite{bishop2011light,wanner2013variational,mitra2012light,Farrugia2017,LFBM5D,GB} followed the traditional paradigm and developed different models for problem formulation. Mitra et al. \cite{mitra2012light} proposed a Gaussian mixture model to encode LF structure for LF spatial SR. Farrugia et al. \cite{Farrugia2017} decomposed HR-LR patches into subspaces and proposed a linear subspace projection method. Alain et al. extended BM3D \cite{BM3D} filtering to LFBM5D for both denoising \cite{alain2017light} and spatial SR \cite{LFBM5D}. Rossi et al. \cite{GB} developed a graph-based method to achieve spatial SR via graph optimization. Although these models can encode the LF structure, their performance is relatively limited due to the poor representation capability of these handcrafted image priors.

Recently, deep CNNs were demonstrated superior to traditional methods in LF spatial SR. In the pioneering work LFCNN \cite{LFCNN15}, SAIs were first super-resolved separately via SRCNN \cite{SRCNN14}, and then fine-tuned in pairs to enhance both spatial and angular resolution. Subsequently, Yuan et al. \cite{LF-DCNN} improved LFCNN by super-resolving each SAI via EDSR \cite{EDSR} and designed an EPI-enhancement network to finetune the initial results. Jin et al. \cite{ATO} proposed an all-to-one method for spatial SR, and performed structural consistency regularization to preserve the parallax structure. Apart from these two-stage methods \cite{LFCNN15,LF-DCNN,ATO}, many one-stage networks have also been proposed for LF spatial SR. Wang et al. \cite{LFNet} proposed a bidirectional recurrent network by extending BRCN \cite{BRCN} to LFs. Zhang et al. \cite{resLF} proposed a multi-stream residual network by stacking SAIs along different angular directions as its inputs. \textcolor{black}{Subsequently, Zhang et al. \cite{MEG-Net} further improved the SR performance by performing 3D convolutions on SAI stacks of different angular directions.} Yeung et al. \cite{LFSSR} proposed LFSSR to alternately reshape LF images between SAI pattern and MacPI pattern for convolution. More recently,  Wang et al. \cite{LF-DFnet} applied deformable convolution to LF images to address the disparity issue for LF spatial SR.
Cheng et al. \cite{ZSLFSR} addressed the domain gap issue in LF spatial SR by applying a zero-shot learning scheme.

\subsection{Angular Super-Resolution}
LF angular SR, also termed as LF reconstruction or view synthesis, aims at reconstructing a densely-sampled LF from a sparse set of views, i.e., super-resolving LFs along the angular dimension. Since a densely-sampled LF can provide smooth parallax shifts and natural refocused details, LF angular SR has been extensively investigated in recent years.

According to the explicit use of depth (or disparity) maps, existing LF angular SR methods can be divided into depth-dependent methods \cite{wanner2013variational,K30,ShearedEPI,shi2020learning,FS-GAF,meng2021light} and depth-independent methods \cite{shi2014light,Shearlet-ASR,LFCNN15,LFEPICNN17,LFEPICNN18,P4DCNN,Yeung2018fast,meng2020high,HDDRNet,SAAN,wu2021revisiting}. Depth-dependent methods achieve LF angular SR by warping and blending input SAIs to target angular positions based on the estimated disparities. Among them, Wanner et al. \cite{wanner2013variational} formulated the depth-based LF reconstruction problem as an energy minimization problem with total viriation regularization. Kalantari et al. \cite{K30} divided angular SR into disparity estimation and color estimation, and developed an end-to-end network to integrate these two stages. Wu et al. \cite{ShearedEPI} implicitly estimated the scene depths by evaluating sheared values of the pre-shifted EPIs, and proposed a CNN for LF reconstruction under large disparity variations. More recently, Shi et al. \cite{shi2020learning} developed a network to adaptively fuse the input images and intermediate features based on the scene depths. Jin et al. \cite{FS-GAF} proposed a flexible LF reconstruction network, which uses a sparsely-sampled unstructure LF to produce a densely-sampled LF with an arbitrary angular resolution. Both disparity and confidence maps were estimated in their method for image warping and view synthesis. Ko et al. \cite{AFR} proposed an adaptive feature remixing approach for spatial and angular SR. In their method, feature of each view was integrated with the ones from adjacent views based on the estimated disparity.

The other branch of methods for LF angular SR is developed without explicit depth estimation. Shi et al. \cite{shi2014light} explored the LF sparsity prior in the Fourier domain to synthesize novel views. Vagharshakyan et al. \cite{Shearlet-ASR} formulated the LF reconstruction problem as an EPI-inpainting problem, and proposed EPI sparse representations in the shearlet transform domain. Learning-based methods have been investigated in recent years. Yoon et al. \cite{LFCNN15} proposed LFCNN for both spatial and angular SR. They first super-resolved each input SAIs individually, and then synthesized novel views sequentially based on their neighboring views. Wu et al. \cite{LFEPICNN17,LFEPICNN18} applied CNNs to 2D EPIs and proposed a blur-restoration-deblur framework for LF reconstruction. Wang et al. \cite{P4DCNN} proposed a pseudo 4D CNN to restore high-frequency details of the EPI volume for angular SR. Yeung et al. \cite{Yeung2018fast} proposed a coarse-to-fine LF reconstruction network by spatial-angular alternate convolution. Meng et al. \cite{HDDRNet} formulated LF spatial and angular SR as a tensor restoration problem and developed a high-dimensional dense residual CNN with 4D convolutions.

\subsection{Disparity Estimation}
LF disparity estimation aims at estimating the spatially-varying displacements among LF images. With the development of LF imaging techniques, many disparity estimation methods have been proposed in recent decades.

Early studies in this area followed the traditional paradigm to perform disparity estimation via geometry analysis and consistency measurement. Tao et al. \cite{tao2013depth} combined the correspondence and defocus cues to estimate the disparity of LFs. Subsequently, shading-based refinement \cite{tao2016depth} and occlusion regularization \cite{wang2016depth} were proposed to improve the performance of \cite{tao2013depth}. Jeon et al. \cite{jeon2015accurate} proposed a phase-based multi-view stereo matching method to estimate sub-pixel shifts in the Fourier domain. Williem et al. \cite{CAE} proposed angular entropy and adaptive defocus costs to handle the noise and occlusion issues for disparity estimation.

Since an EPI contains patterns of oriented lines and the slope of these lines is related to the disparity values, many methods achieve disparity estimation by analyzing the slope of each line on EPIs. Wanner et al. \cite{wanner2013variational} proposed a structure tensor to estimate the slope of lines on horizontal and vertical EPIs, and refined the initial results by global optimization. Zhang et al. \cite{SPO} proposed a spinning parallelogram operator (SPO) to estimate the slopes for disparity estimation. Sheng et al. \cite{SPO-MO} proposed to estimate slopes using multi-orientation EPIs and achieved improved results over SPO. Schilling et al. \cite{OBER} presented an inline occlusion handling scheme operated on EPIs to achieve state-of-the-art performance among traditional methods.

Recently, deep CNNs have been widely used for disparity estimation and achieved superior performance over traditional methods. Heber et al. \cite{heber2016convolutional} proposed a CNN to learn an end-to-end mapping between a 4D LF and its corresponding depths. Subsequently, Heber et al. \cite{heber2017neural} proposed a U-Net with 3D convolutions to extract geometric information for robust disparity estimation. Shin et al. \cite{EPINET} proposed a multi-stream network (i.e., EPINET) and a specifically designed data augmentation approach for fast and accurate disparity estimation. EPINET significantly improves the accuracy of disparity estimation and achieves top performance on the HCInew benchmark \cite{HCInew} at that time.  Tsai et al. \cite{LFAttNet} proposed an attention-based view selection network (i.e., LFattNet) to adaptively incorporate all angular views for disparity estimation. Chen et al. \cite{AttMLFNet} proposed an attention-based multi-level fusion network to handle the occlusion problem for disparity estimation.

\subsection{CNN Architectures for LF Image Processing}
In this subsection, we briefly review several widely-used CNN architectures for LF image processing.

\textbf{1) Neighbor-view combination.} The pioneering CNN-based LFSR method LFCNN \cite{LFCNN15,LFCNN17} was developed on the neighbor-view combination framework. In their methods, SAIs are first processed independently using SRCNN \cite{SRCNN14}, and then fine-tuned in pairs or quads to incorporate angular information. Since only 2 or 4 adjacent views were used to super-resolve$/$reconstruct a specific view, the neighbor-view based method cannot achieve a high reconstruction quality due to the discard of rich angular information.

\textbf{2) Convolution on EPIs.} Since an EPI can encode LF structures into regular line patterns with different slopes, several methods \cite{LFEPICNN17,LFEPICNN18,ShearedEPI,SAAN,ORM} have been proposed to process LF images by performing convolutions on EPIs. Compared to neighbor-view based methods which only incorporate angular information from adjacent views, EPI-based methods can exploit angular information from more views and thus achieve improved performance. However, since an EPI is a 2D slice of a 4D LF, performing convolutions on EPIs can only incorporate angular information from the same horizontal or vertical views and cannot incorporate the spatial context prior. The underuse of spatial and angular information limits the performance of EPI-based methods, especially on complex scenes.

\textbf{3) Multi-stream structure.} In multi-stream networks \cite{EPINET,resLF,MALFRNet,AttMLFNet}, SAIs are first stacked along four angular directions (i.e., horizontal, vertical, diagonal, and anti-diagonal) to form the inputs of each stream. Then, features generated by different streams are integrated and fed to the merge module for final prediction \cite{EPINET,EPI-Shift} or reconstruction \cite{resLF,MALFRNet}. Compared to neighbor-view based and EPI-based networks, multi-stream networks can exploit information from more angular views. However, multi-stream networks discard the views outside the four angular directions, resulting in under-exploitation of the rich angular information in an LF (e.g., 33 views are used in a 9$\times$9 LF).

\textbf{4) Spatial-angular alternate convolution.} Yeung et al. \cite{LFSSR,Yeung2018fast} proposed spatial-angular alternate convolutions to process all angular views in one forward pass. Specifically, LF features are alternately reshaped to a stack of SAIs or a stack of macro-pixels, 3$\times$3 convolutions are then performed on the features at each stage. By cascading the reshape and convolution operation, the network achieves promising performance on LF reconstruction \cite{Yeung2018fast} and LF image SR \cite{LFSSR}. This scheme was then adopted in several LF image processing methods \cite{ATO,guo2020deep,LFASR-geo,guo2021deep} as a regularization term to embed the LF structural prior. However, these networks can only process information within a single domain at each stage, and thus cannot exploit deeper representation in neither spatial nor angular dimension.

Apart from these aforementioned networks, many architectures have also been developed for LF image processing, such as bi-directional recurrent structure \cite{LFNet}, 4D convolution \cite{HDDRNet,meng2020high,Mask4D} and cost volumes \cite{LFAttNet,AttMLFNet}. The key difference between existing schemes and our disentangling mechanism is, we can fully use the information from all angular views and incorporate the LF structure prior. With our mechanism, high-dimensional LF data can be disentangled into different low-dimensional subspaces. Consequently, the difficulty for learning deep CNNs is reduced and several LF image processing tasks can be benefited.

\section{The LF Disentangling Mechanism}\label{sec:DistgMechanism}
In this section, we present our disentangling mechanism in details. We first introduce the LF representation, then describe our convolutions for LF feature disentanglement.

 \subsection{Light Field Representation}

 \textcolor{black}{We use the two-plane model \cite{levoy1996light} to parameterize LFs. That is, an LF can be formulated as a 4D tensor $\mathcal{L}(u,v,h,w)\in\mathbb{R^{\mathrm{\mathit{U}\times\mathit{V}\times\mathit{H}\times\mathit{W}}}}$, where $\mathit{U}$ and $\mathit{V}$ represent the angular dimensions, and $\mathit{H}$ and $\mathit{W}$ represent the spatial dimensions. Since 4D tensors cannot be directly visualized via 2D images, LFs are generally organized into the following forms for visualization.}

 \begin{figure}
 \centering
 \includegraphics[width=8.8cm]{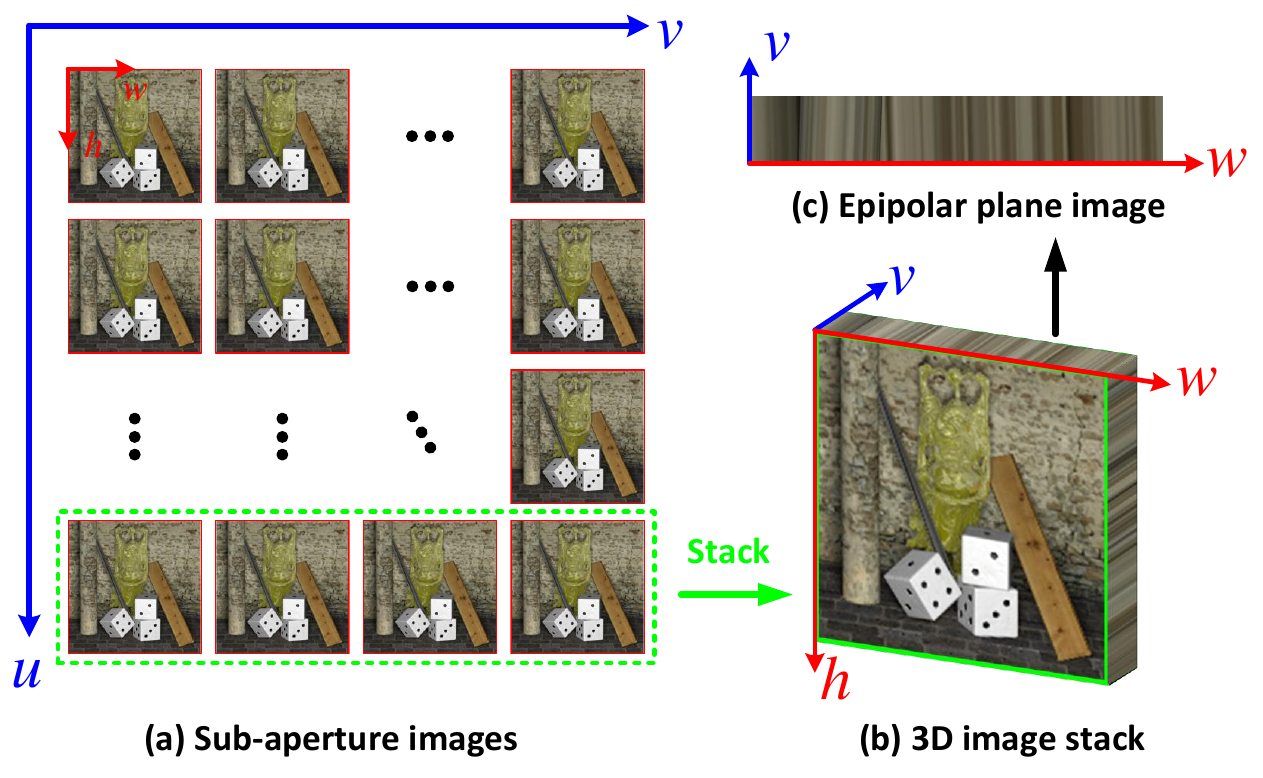}
 \caption{The SAI and EPI representation of 4D LFs.}
 \label{figLFrepresentation}
 \end{figure}

\subsubsection{Sub-aperture Image (SAI)}
 As shown in Fig.~\ref{figLFrepresentation}(a), a 4D LF can be organized as a $U\times V$ array of SAIs of size $H\times W$. These SAIs can be considered as the images recorded by different cameras on the angular plane, and the SAI at angular coordinate $(u,v)$ can be denoted as $\mathcal{L}\left(u,v,:,:\right)\in\mathbb{R^{\mathrm{\mathit{H}\times\mathit{W}}}}$. SAIs have similar styles as 2D natural images and can facilitate spatial information extraction. However, when an LF is organized as an array of SAIs, the angular information is implicitly contained among different SAIs and thus is difficult to extract.

\subsubsection{ Epipolar Plane Image (EPI)}
 As shown in Figs.~\ref{figLFrepresentation}(b) and \ref{figLFrepresentation}(c), if SAIs are stacked along the angular dimension $V$, an EPI $\mathcal{L}\left(u,:,h,:\right)\in\mathbb{R^{\mathrm{\mathit{V}\times\mathit{W}}}}$ can be obtained by viewing this 3D image stack on the spatial-angular plane. Since objects in space can be projected onto different spatial locations on different angular views, an EPI contains patterns of oriented lines whose slopes reflect the disparity values. Consequently, EPIs are widely used to infer 3D scene geometries for LF reconstruction \cite{LFEPICNN17, LFEPICNN18, ShearedEPI, SAAN} and depth estimation \cite{wanner2013variational, SPO, SPO-MO, OBER, ORM}. However, since an EPI is only a 2D horizontal or vertical slice of a 4D LF, it is difficult to encode all the spatial and angular cues via EPIs, resulting in the under-exploitation of useful information.

\begin{figure}
\centering
\includegraphics[width=8.8cm]{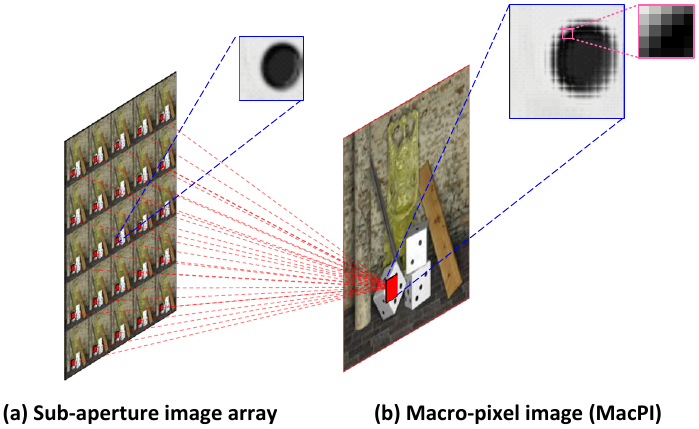}
\caption{An illustration of the relationship between SAI and MacPI representations. To convert an SAI array into a MacPI, pixels at the same spatial locations of each SAI need to be organized according to their angular coordinates to generate a macro-pixel \textcolor{black}{(the 5$\times$5 patch on the top-right corner of (b))}. Then, the generated macro-pixels need to be organized according to their spatial coordinates. See details in the \href{https://yingqianwang.github.io/DistgLF/Appendix.pdf}{Appendix}.}
\label{figSAIMacPI}
\end{figure}

  \subsubsection{ Macro-pixel Image (MacPI)}
  Similar to SAI arrays, a 4D LF can be also organized as an $H\times W$ array of macro-pixels. A macro-pixel  $\mathcal{L}\left(:,:,h,w\right)\in\mathbb{R^{\mathrm{\mathit{U}\times\mathit{V}}}}$ can be considered as a set of pixels with same spatial coordinate $(h,w)$ but captured by cameras at different angular locations\footnote{\textcolor{black}{In the area of LF disparity estimation, ``macro-pixels'' is termed as ``angular patches under zero disparity''. In this paper, we keep the terminology consistent to our previous conference paper \cite{LF-InterNet}, and use ``macro-pixel image (MacPI)'' to denote the image obtained by organizing macro-pixels according to their spatial coordinates.}}. As shown in Fig.~\ref{figSAIMacPI}, to convert an SAI array into a MacPI, pixels at the same spatial locations of each SAI need to be organized according to their angular locations to generate a macro-pixel. Then, the generated macro-pixels need to be organized according to their spatial coordinates. Details of the SAI-MacPI conversion are introduced in the \href{https://yingqianwang.github.io/DistgLF/Appendix.pdf}{Appendix}. Note that, due to the ``mosaic pattern'' introduced by macro-pixels (see the zoom-in region in Fig.~\ref{figSAIMacPI}(b)), MacPIs are unfriendly to human visual perception. However, with the MacPI representation, the spatial and angular information in an LF is evenly mixed. In this case, it is convenient to extract and incorporate spatial and angular information using convolutions. Therefore, in this paper, we use the MacPI representation and design a class of convolutions for feature extraction and disentanglement, which will be introduced in the next subsection.

 \subsection{LF Feature Disentanglement}\label{sec:FeatureExtractors}
\begin{figure}
\centering
\includegraphics[width=8.8cm]{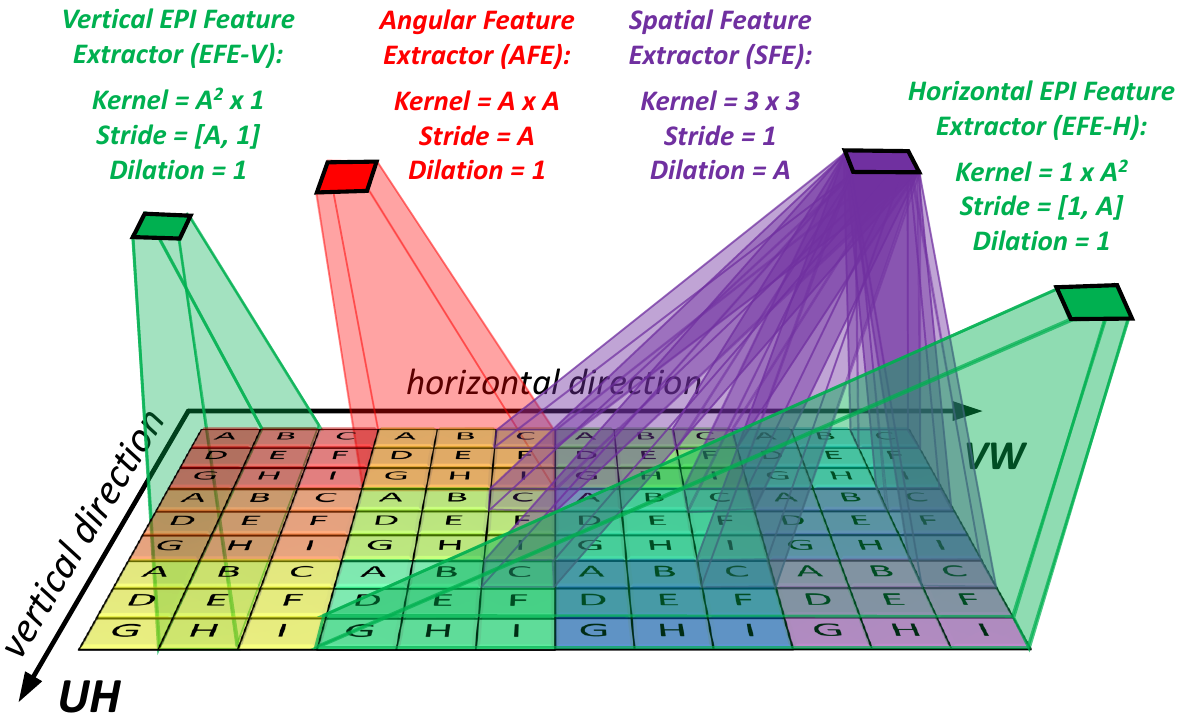}
\caption{{An illustration of the spatial, angular, and EPI feature extractors. \textcolor{black}{Here, an LF of size $U$$=$$V$$=$$3$ (i.e., $A$$=$$3$), $H$$=$$3$, $W$$=$$4$ is used as a toy example.} For better visualization of the MacPI, different macro-pixels are paint with different background colors while pixels from different views are denoted with different characters. The proposed feature extractors can disentangle LFs into different subspaces, i.e., an SFE convolves pixels from the same views, an AFE convolves pixels from the same macro-pixel, and an EFE convolves pixels on EPIs.}
\label{fig:FE}}
\end{figure}

  We take the 4D LF structure into consideration and design three types of feature extractors for feature extraction. By applying the designed feature extractors to MacPIs, the 4D LF can be disentangled into different 2D subspaces. Here, we use a toy example (see Fig.~\ref{fig:FE}) to illustrate the proposed feature extractors. Note that, since most LF image processing methods use SAIs distributed in a square array as their inputs, in this paper, we set $U$$=$$V$$=$$A$, where $A$ denotes the angular resolution.

\subsubsection{Spatial Feature Extractor}
 \textcolor{black}{To extract spatial features from a MacPI, pixels from the same view (i.e., with same angular coordinates) should be convolved while pixels from different views should be isolated. To achieve this goal, we design our spatial feature extractor (SFE) as a convolution with a kernel size of 3$\times$3, a stride of 1, and a dilation of $\mathit{A}$. Zero padding is performed to ensure that the output has the same spatial size as the input MacPI. Here, we follow most existing networks to use 3$\times$3 convolutions to achieve a relatively small model size, and use a dilation of $\mathit{A}$ to isolate information from different views on a MacPI. As a result, when applying our SFE to MacPI, only spatial information in the LF is processed.}

 \subsubsection{Angular Feature Extractor}
 \textcolor{black}{To extract angular features from a MacPI, pixels in a macro-pixel (i.e., with same spatial coordinates) should be convolved while pixels from different macro-pixels should not be aliased. To achieve this goal, we design our angular feature extractor (AFE) as a convolution with a kernel size of $A$$\times$$A$ and a stride of $A$. Padding is not performed so that the output has a spatial size of $H$$\times$$W$. Here, the $A$$\times$$A$ kernel can make our AFE convolve pixels in a macro-pixel, and the stride of $A$ can ensure pixels from different macro-pixels not be aliased. As a result, when applying our AFE to MacPI, only angular information of the LF is processed.}

 \subsubsection{EPI Feature Extractor}
 Although SFE and AFE can disentangle LFs into spatial and angular subspaces, the relationship between the spatial and angular cues is overlooked. Since the line patterns on EPIs can well reflect the spatial-angular correlation, in this paper, we design horizontal and vertical EPI feature extractors (i.e., EFE-H and EFE-V) to disentangle LFs into $V\text{-}W$ and $U\text{-}H$ subspaces, respectively. Different from SFE and AFE, EFEs have asymmetric kernels and strides. Without loss of generality, we take the EFE-H as an example to introduce its definition and implementation.

 \textcolor{black}{As shown in Fig.~\ref{fig:FE}, to extract features in the $V\text{-}W$ subspace, pixels at a horizontal slice (i.e., with the same $U$ and $H$ coordinates) on the MacPI should be convolved. To achieve this goal, we design our EFE-H as a convolution with a kernel size of 1$\times$$A$$^2$, a vertical stride of 1 and a horizontal stride of $A$. Padding is not performed so that the output feature has a spatial size of $AH$$\times$$W$. Here, the 1$\times$$A$$^2$ kernel and the vertical stride can make our EFE-H convolve pixels on each horizontal slice, and the horizontal stride can keep the LF structure during convolution.}
 According to the 4D LF structure, applying an EFE-H to MacPIs is equivalent to performing an $A$$\times$$A$ convolution with a stride of 1 on the horizontal EPIs (i.e., the $V\text{-}W$ slice of the LF). Consequently, EFE-H can simultaneously process information from one spatial dimension (i.e., dimension $W$) and one angular dimension (i.e., dimension $V$), and thus is powerful in spatial-angular correlation modeling.

 \subsubsection{\textcolor{black}{Discussion}}
 \textcolor{black}{The aforementioned feature extractors can fully incorporate the LF structure prior and disentangle LFs into different subspaces. Since each feature extractor only processes LF features within a 2D subspace, the complexity of the 4D LF data is effectively reduced, making LF representation learning much easier. In practice, we combine these feature extractors into different modules for different tasks. By stacking multiple modules (i.e., deepening the networks), our feature extractors can work jointly to incorporate the information from different domain, and enlarge the receptive field to cover the spatially varying disparities. In the following sections, we apply our disentangling mechanism to three typical LF image processing tasks, and introduce the details of our modules and networks.}

 \begin{figure*}[t]
 \centering
 \includegraphics[width=17.5cm]{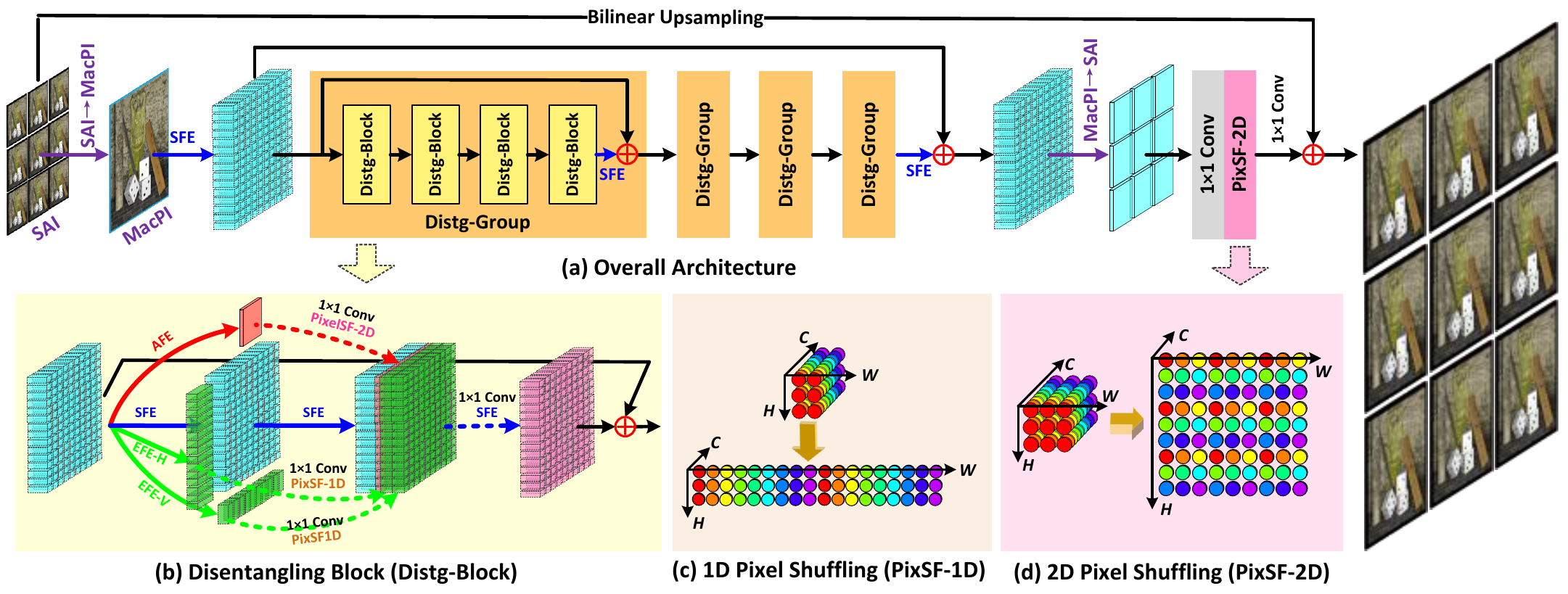}
 \caption{An overview of our DistgSSR network.}\label{fig:DistgSSR}
 \end{figure*}

 \section{DistgSSR: Disentangling Mechanism for Spatial Super-Resolution}\label{sec:DistgSSR}
 In this section, we apply our disentangling mechanism to LF spatial SR and propose a network named DistgSSR. The network design and experimental results are presented in the following subsections.

 \subsection{Network Design}
 \subsubsection{Overview}
 An overview of our DistgSSR is shown in Fig.~\ref{fig:DistgSSR}(a). Given an LR SAI array $\mathcal{I^{LR}_{\mathit{SAIs}}}\in\mathbb{R^{\mathrm{\mathit{AH}\times\mathit{AW}}}}$, we first convert it into an LR MacPI $\mathcal{I^{LR}_{\mathit{MacPI}}}\in\mathbb{R^{\mathrm{\mathit{AH}\times\mathit{AW}}}}$, and then use the aforementioned spatial, angular, and EPI convolutions (Section \ref{sec:FeatureExtractors}) to disentangle and process features in the MacPI pattern. Following RCAN \cite{RCAN}, we develop our DistgSSR based on the residual-in-residual structure for better SR performance. Specifically, four residual disentangling groups are used in our DistgSSR, with each group comprising of four residual disentangling blocks. The final output of our network is an HR SAI array $\mathcal{I^{HR}_{\mathit{SAIs}}}\in\mathbb{R^{\mathrm{\mathit{\alpha AH}\times\mathit{\alpha AW}}}}$, where $\alpha$ denotes the upscaling factor. \textcolor{black}{We follow \cite{resLF, ATO, LF-DFnet} to convert input images into the YCbCr color space, and only super-resolve the Y channel of images, leaving Cb and Cr channel images being bicubicly upscaled.}

 \subsubsection{Distg-Block for SSR}
 The basic module of our DistgSSR is the disentangling block (Distg-Block). As shown in Fig.~\ref{fig:DistgSSR}(b), the Distg-Block takes a MacPI feature $\mathcal{F_{\mathit{in}}}\in\mathbb{R^{\mathrm{\mathit{AH}\times\mathit{AW}\times\mathit{C}}}}$ \textcolor{black}{($C$ is the number of channels and is set to 64 in our DistgSSR)} as its inputs and uses four parallel branches (i.e., an angular branch, a spatial branch, and two EPI branches) to achieve feature disentanglement and incorporation. Specifically, in the spatial branch, two SFEs\footnote{All convolutions except for the last SFE in our Distg-Block are followed with a LeakyReLU layer (with a leaky factor of 0.1) for activation.} are sequentially performed to incorporate the spatial information within each view. Note that, features in this branch have identical number of channels as the input feature. Finally, the spatial branch produces an output feature $\mathcal{F_{\mathit{spa}}}\in\mathbb{R^{\mathit{AH\times AW\times C}}}$.

 In the angular branch, an AFE is first performed to produce an intermediate angular feature $\mathcal{F^{\mathit{tmp}}_{\mathit{ang}}}\in\mathbb{R^{\mathit{H\times W\times \frac{C}{4}}}}$. Then, this intermediate feature is upsampled by a factor of $A$ to generate the output feature $\mathcal{F_{\mathit{ang}}}\in\mathbb{R^{\mathit{AH\times AW\times \frac{C}{4}}}}$. Since pixels in a macro-pixel can be unevenly distributed due to edges and occlusions in real scenes \cite{park2017robust}, we learn this discontinuity using a $1\times1$ convolution and a 2D pixel shuffling layer (as shown in Fig.~\ref{fig:DistgSSR}(d)) for angular-to-spatial upsampling.

 Although the aforementioned spatial and angular branches can disentangle and process LF features in the spatial and angular domain, the disparity issue among different views has not been well addressed. Specifically, due to the 3D property of real-world scenes, objects at different depths have different disparity values. Therefore, pixels of an object among different views cannot always locate at a single macro-pixel on the MacPI features. This disparity issue impedes AFE to incorporate angular information effectively and can lead to performance degradation when handling LFs with large disparity variations. Since EPIs can encode disparities into lines with different slopes, we introduce two orthogonal EPI branches with shared weights to extract features on the ``\textit{U-H}'' and ``\textit{V-W}'' subspaces to achieve disparity-robust information incorporation. Specifically, the input MacPI feature are fed to EFE-H and EFE-V to generate intermediate EPI feature $\mathcal{F^{\mathit{tmp}}_{\mathit{epih}}}\in\mathbb{R^{\mathit{AH\times W\times \frac{C}{2}}}}$ and $\mathcal{F^{\mathit{tmp}}_{\mathit{epiv}}}\in\mathbb{R^{\mathit{H\times AW\times \frac{C}{2}}}}$. Then, these two intermediate features are upsampled along different directions to produce output features $\mathcal{F_{\mathit{epih}}}\in\mathbb{R^{\mathit{AH\times AW\times \frac{C}{2}}}}$ and $\mathcal{F_{\mathit{epiv}}}\in\mathbb{R^{\mathit{AH\times AW\times \frac{C}{2}}}}$. Similar to the angular branch, we use a $1\times1$ convolution and a 1D pixel shuffling layer (as shown in Fig.~\ref{fig:DistgSSR}(c)) for 1D upsampling.

 Finally, features generated by different branches (i.e., $\mathcal{F_{\mathit{spa}}}$, $\mathcal{F_{\mathit{ang}}}$, $\mathcal{F_{\mathit{epih}}}$, $\mathcal{F_{\mathit{epiv}}}$) are concatenated and fused by a 1$\times$1 convolution and an SFE. In this way, the complementary angular information can be well used to guide spatial information extraction. Note that, the fused feature $\mathcal{F_{\mathit{fused}}}$ are added with the input feature $\mathcal{F_{\mathit{in}}}$ to achieve local residual learning.

\subsubsection{Spatial Upsampling}
To increase the spatial resolution of the LF features, we first reshape the feature generated by the cascaded Distg-Groups from the MacPI pattern to the SAI pattern (see \href{https://yingqianwang.github.io/DistgLF/Appendix.pdf}{Appendix} for details), then use a $1\times1$ convolution to increase their depth to $\alpha^2 C$ (where $\alpha$ is the upsclaing factor). A 2D pixel shuffling layer is used to upsample the features to the target resolution $\alpha AH \times \alpha AW$. Finally, a $1\times1$ convolution is applied to squeeze the number of channels to 1 to generate super-resolved SAIs.

\begin{table}
\caption{PSNR results achieved by several variants of our DistgSSR for $2\times$SR. Note that, the number of parameters (i.e., \textbf{\textit{\#Param.}}) of different variants were adjusted to make their model sizes not smaller than our DistgSSR. }\label{tab:ablationSSR}
\renewcommand\arraystretch{1.2}
\centering
\scriptsize
\begin{tabular}{|p{0.05cm}<{\centering} | p{0.11cm}<{\centering} p{0.11cm}<{\centering} p{0.17cm}<{\centering} | p{0.6cm}<{\centering} | p{0.45cm}<{\centering} p{0.55cm}<{\centering} p{0.55cm}<{\centering} p{0.45cm}<{\centering} p{0.48cm}<{\centering}|p{0.55cm}<{\centering}|}
\hline
    & \tiny{\textbf{\textit{Spa}}} &  \tiny{\textbf{\textit{Ang}}} &  \tiny{\textbf{\textit{EPI}}}  & \tiny{\textbf{\textit{\#Param.}}}& \tiny{\textbf{\textit{EPFL}}} &  \tiny{\textbf{\textit{HCInew}}} &  \tiny{\textbf{\textit{HCIold}}} &  \tiny{\textbf{\textit{INRIA}}} &  \tiny{\textbf{\textit{STFgtr}}}& \tiny{\textbf{\textit{Average}}}\\
\hline
 1 & $\checkmark$  & & & 3.59M & 33.04 & 34.93 & 41.15 & 34.97 & 36.56 & 36.13 \\
 2 &  & $\checkmark$  &   & 3.55M &29.58 & 31.28 & 36.52 & 30.90 & 30.05 & 31.66 \\
 3 &$\checkmark$  & $\checkmark$  &  & 3.62M   &34.48 & 37.65 & 44.74 & 36.26 &39.96 &38.62 \\
 4 & $\checkmark$  & $\checkmark$  & $\checkmark^{\star}$  & 4.43M  & 34.53 & 37.86 & 44.81 & 36.25 & 40.33& 38.76 \\
\hline
\end{tabular}
\vspace{0.05cm}

\begin{tabular}{|p{1.66cm}<{\centering}|p{0.6cm}<{\centering}|p{0.45cm}<{\centering} p{0.55cm}<{\centering} p{0.55cm}<{\centering} p{0.45cm}<{\centering} p{0.48cm}<{\centering}|p{0.55cm}<{\centering}|}
\hline
 \textit{4D Convolution} & 22.3M & 34.59 & 37.86 & 44.68 & 36.42 & 39.50 & 38.61 \\
 \textit{DistgSSR} & \textbf{3.53M} & \textbf{34.80} & \textbf{37.95} & \textbf{44.92} & \textbf{36.58} & \textbf{40.37} & \textbf{38.92} \\
\hline
\end{tabular}
\leftline{$^{\star}$Weight sharing is not performed between the two EPI branches in \textit{model-4}.}
\end{table}

\subsection{Experiments}
In this section, we first introduce the datasets and our implementation details. Then, we conduct ablation studies to investigate our design choices. Finally, we compare our network to several state-of-the-art SISR and LF image SR methods.

\subsubsection{Datasets and Implementation Details}
We followed \cite{LF-DFnet} to use 5 public LF datasets (i.e., EPFL \cite{EPFL}, HCInew \cite{HCInew}, HCIold \cite{HCIold}, INRIA \cite{INRIA}, STFgantry \cite{STFgantry}) for both training and test. The division of training and test set was set identical to that in \cite{LF-DFnet}. All LFs in these datasets have an angular resolution of $9\times9$. In the training stage, we cropped each SAI into patches with a stride of 32, and used the bicubic downsapling approach to generate LF patches of size $32\times32$. We performed random horizontal flipping, vertical flipping, and 90-degree rotation to augment the training data by 8 times. Note that, the spatial and angular dimension need to be flipped or rotated jointly to maintain LF structures.

Our network was trained using the L1 loss and optimized using the Adam method \cite{Adam} with $\beta_1$=0.9, $\beta_2$=0.999 and a batch size of 8. Our DistgSSR was implemented in PyTorch on a PC with two NVidia RTX 2080Ti GPUs. The learning rate was initially set to $2\times10^{-4}$ and decreased by a factor of 0.5 for every 15 epochs. The training was stopped after 50 epochs.

\textcolor{black}{Following \cite{EDSR,RCAN,resLF,LF-DFnet}, we used PSNR and SSIM calculated on the Y channel images as quantitative metrics for performance evaluation.} Note that, to obtain the metric score (e.g., PSNR) for a dataset with $M$ test scenes (each scene with an angular resolution of $A\times A$), we first calculated the metric on $A\times A$ SAIs on each scene separately, then obtained the score for each scene by averaging its $A^2$ scores, and finally obtained the score for this dataset by averaging the scores of all $M$ scenes.

\subsubsection{Ablation Study}
To investigate the benefits introduced by different design choices, in this subsection, we compare the performance of our DistgSSR with different architectures and angular resolutions.

\textbf{1) Branches in the Distg-Block.}
We demonstrate the effectiveness of the spatial, angular and EPI branches in our Distg-Block by selectively removing them from our network. Note that, the channel numbers of all the variants were adjusted to make their model sizes not smaller than our main model. Table~\ref{tab:ablationSSR} shows the comparative PSNR results.

 \textit{Spatial Only}: We introduced \textit{model-1} by using the spatial branch only for SR. Without using angular and EPI branches, \textit{model-1} cannot incorporate any angular information and is identical to an SISR network. As shown in Table~\ref{tab:ablationSSR}, \textit{model-1} suffers a decrease of 2.79dB in PSNR as compared to DistgSSR (36.13 v.s. 38.92), which demonstrates the importance of angular information for LF spatial SR.

 \textit{Angular Only}: We introduced \textit{model-2} by using the angular branch only for SR. Consequently,  \textit{model-2} cannot incorporate any spatial information and thus achieves significantly lower PSNR values than DistgSSR. It demonstrates that the spatial information is crucial for LF spatial SR while the angular information can further be used as a complementary part to spatial information.

 \textit{Spatial$+$Angular}: We validate the effectiveness of EPI branches by removing them from our DistgSSR. Without using EPI branches, \textit{model-3} suffers a decrease of 0.30dB in PSNR as compared to DistgSSR (38.62 v.s. 38.92). Note that, the PSNR drop is more significant on the datasets with large disparity variations (e.g., 0.41 dB PSNR drop on the STFgantry \cite{STFgantry} dataset). That is because, the proposed angular feature extractor cannot effectively incorporate angular information under large disparity variations since pixels of an object among different views can locate at different macro-pixels. In contrast, the EPI feature extractor can disentangle LFs into EPI subspaces, and its powerful spatial-angular correlation modeling capability makes our DistgSSR more robust to disparity variations.

 \textit{EPI branch w/o weight sharing}: We introduced \textit{model-4} by removing the weight sharing between horizontal and vertical EPI branches. As shown in Table~\ref{tab:ablationSSR}, the performance of \textit{model-4} are slightly inferior to the main model. That is because, the horizontal and vertical EPI slices share similar patterns. By performing weight sharing between horizontal and vertical EPI branches, our network can be better regularized to fit the LF parallax structure.

\textbf{2) Distg-Block v.s. 4D Convolution.} Since several recent works \cite{HDDRNet,meng2020high,Mask4D} used 4D convolutions to handle LF data and have achieved promising performance, we compare our disentangling mechanism with 4D convolutions by replacing our Distg-Block with a series of 4D residual blocks. As shown in Table~\ref{tab:ablationSSR}, stacking 4D convolutions can result in a very large model size (i.e., 22.3M for 2$\times$SR) but cannot introduce performance improvements. By using 4D convolutions, the variant achieves an average PSNR of 38.61dB for 2$\times$SR, which is 0.31dB lower than our DistgSSR. That is because, by using our proposed Distg-Block, the high-dimensional LF data can be disentangled into different subspaces and the inherent structural characteristics of input LF images can be efficiently learned by our specifically designed feature extractors.

\begin{table}
\caption{PSNR results achieved by our DistgSSR with different angular resolutions for $2\times$SR.} \label{tab:angRes}
\centering
\renewcommand\arraystretch{1.2}
\scriptsize
\begin{tabular}{|c|ccccc|c|}
\hline
  \textbf{\textit{AngRes}} & 
  \textbf{\textit{EPFL}} &  \textbf{\textit{HCInew}} & \textbf{\textit{HCIold}} &  \textbf{\textit{INRIA}} &  \textbf{\textit{STFgtr}} & \textbf{\textit{Average}}\\
\hline
  2$\times$2 &  33.71 & 36.76 & 43.62 & 35.67 & 38.96 & 37.74 \\
  3$\times$3 &  34.14 & 37.40 & 44.34 & 35.97 & 39.79 & 38.33 \\
  4$\times$4 &  34.41 & 37.75 & 44.75 & 36.34 & 40.27 & 38.70 \\
  5$\times$5 &  34.80 & 37.95 & 44.92 & 36.58 & 40.37 & 38.92 \\
  6$\times$6 &  35.00 & 38.00 & 44.95 & 36.63 & 40.55 & 39.03 \\
  7$\times$7 &  35.16 & 38.10 & 44.94 & 36.68 & 40.57 & 39.09 \\
  8$\times$8 &  34.95 & \textbf{38.23} & 45.06 & 36.62 & \textbf{40.78} & 39.13 \\
  9$\times$9 &  \textbf{35.23} & 38.15 & \textbf{45.12} & \textbf{36.70} &  40.70 & \textbf{39.18} \\
\hline
\end{tabular}
\end{table}

\begin{table*}
\caption{
PSNR$/$SSIM values achieved by different methods for $2\times$ and $4\times$SR. The best results are in \textbf{bold faces}.} \label{tab:quantitativeSSR}
\centering
\scriptsize
\renewcommand\arraystretch{1.25}
\begin{tabular}{|l|p{1.15cm}<{\centering} p{1.15cm}<{\centering} p{1.15cm}<{\centering} p{1.15cm}<{\centering} p{1.15cm}<{\centering}| p{1.15cm}<{\centering} p{1.15cm}<{\centering} p{1.15cm}<{\centering} p{1.15cm}<{\centering} p{1.15cm}<{\centering}|}
\hline
\multirow{2}*{Method}&   \multicolumn{5}{c|}{$2\times$}  &   \multicolumn{5}{c|}{$4\times$} \\
\cline{2-11}
  & EPFL &HCInew & HCIold & INRIA & STFgantry  &   EPFL &HCInew & HCIold & INRIA & STFgantry \\
\hline
Bicubic                     & 29.50/0.935 & 31.69/0.934 & 37.46/0.978 & 31.10/0.956 & 30.82/0.947 & 25.14/0.831 & 27.61/0.851 & 32.42/0.934 & 26.82/0.886 & 25.93/0.843 \\
VDSR \cite{VDSR}  & 32.50/0.960 & 34.37/0.956 & 40.61/0.987 & 34.43/0.974 & 35.54/0.979 & 27.25/0.878 & 29.31/0.883 & 34.81/0.952 & 29.19/0.921 & 28.51/0.901 \\
EDSR \cite{EDSR}   & 33.09/0.963 & 34.83/0.960 & 41.01/0.988 & 34.97/0.977 & 36.29/0.982 & 27.84/0.886 & 29.60/0.887 & 35.18/0.954 & 29.66/0.926 & 28.70/0.908 \\
RCAN \cite{RCAN} & 33.16/0.964 & 34.98/0.960 & 41.05/0.988 & 35.01/0.977 & 36.33/0.983 & 27.88/0.886 & 29.63/0.888 & 35.20/0.954 & 29.76/0.927 & 28.90/0.911 \\
\hline
LFBM5D \cite{LFBM5D}  & 31.15/0.955 & 33.72/0.955 & 39.62/0.985 & 32.85/0.970 & 33.55/0.972 & 26.61/0.870 & 29.13/0.882 & 34.23/0.951 & 28.49/0.914 & 28.30/0.900 \\
GB \cite{GB}                    & 31.22/0.959 & 35.25/0.969 & 40.21/0.988 & 32.76/0.972 & 35.44/0.984 & 26.02/0.863 & 28.92/0.884 & 33.74/0.950 & 27.73/0.909 & 28.11/0.901 \\
resLF \cite{resLF}             & 33.62/0.971 & 36.69/0.974 & 43.42/0.993 & 35.39/0.981 & 38.36/0.990 & 28.27/0.904 & 30.73/0.911 & 36.71/0.968 & 30.34/0.941 & 30.19/0.937 \\
LFSSR \cite{LFSSR}        & 33.69/0.975 & 36.86/0.975 & 43.75/{0.994} & 35.27/0.983 & 38.07/0.990 & 28.27/0.908 & 30.72/0.912 & 36.70/0.969 & 30.31/0.945 & 30.15/0.939 \\
LF-ATO \cite{ATO}               & 34.27/0.976 & 37.24/0.977 & 44.20/{0.994} & 36.15/0.984 & {39.64}/0.993 & 28.52/0.912 & 30.88/0.914 & 37.00/0.970 & 30.71/0.949 & 30.61/0.943 \\
LF-InterNet \cite{LF-InterNet} & 34.14/0.976 & 37.28/0.977 & {44.45}/\textbf{0.995} & 35.80/{0.985} & 38.72/0.992 & 28.67/0.914 & 30.98/0.917 & 37.11/{0.972} & 30.64/0.949 & 30.53/0.943 \\
LF-DFnet \cite{LF-DFnet}       & {34.44/0.977} & {37.44/0.979} & 44.23/{0.994} & {36.36}/0.984 & 39.61/{0.994} & {28.77/0.917} & {31.23/0.920} & {37.32/0.972} & {30.83/0.950} & {31.15/0.949} \\
DistgSSR (ours)  & \textbf{34.80}/\textbf{0.979} & \textbf{37.95}/\textbf{0.980} & \textbf{44.92}/\textbf{0.995} & \textbf{36.58}/\textbf{0.986} & \textbf{40.37}/\textbf{0.994} & \textbf{28.98}/\textbf{0.919} & \textbf{31.38}/\textbf{0.922} & \textbf{37.55}/\textbf{0.973} & \textbf{30.99}/\textbf{0.952} & \textbf{31.63}/\textbf{0.953} \\
\hline
\end{tabular}
\end{table*}

\begin{figure*}
\centering
\includegraphics[width=18cm]{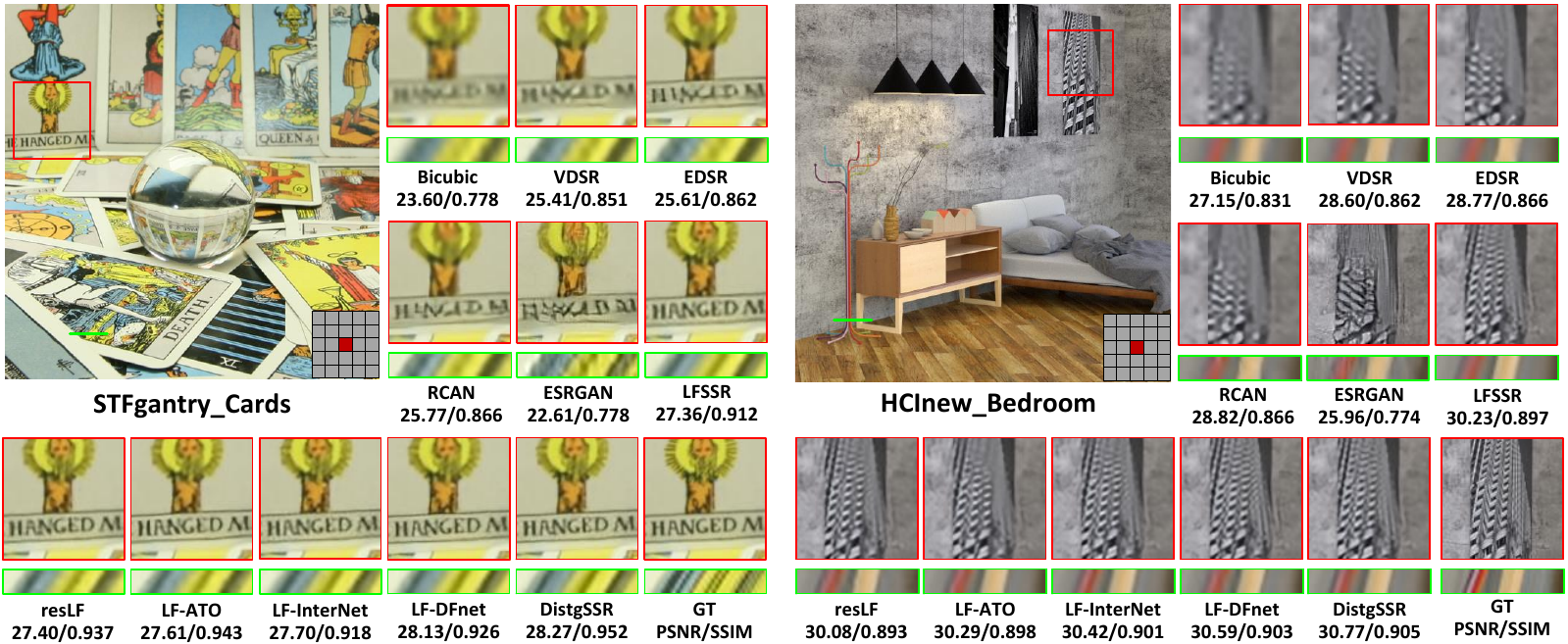}
\caption{\textcolor{black}{Visual comparisons for 4$\times$SR.} The super-resolved center view images and horizontal EPIs are shown. The PSNR and SSIM scores achieved by different methods on the presented scenes are reported below the zoom-in regions.}
\label{fig:Visual-SSR}
\end{figure*}

\textbf{3) Angular resolution.} We analyze the performance of DistgSSR with different angular resolutions. Specifically, we extract the central $A$$\times$$A$ ($A$=2, 3, $\cdots$, 9) SAIs from the input LFs, and trained different models for 2$\times$SR. As shown in Table~\ref{tab:angRes}, the PSNR values are improved as the angular resolution is increased. That is because, additional views can provide rich angular information which is beneficial to spatial SR. It is also notable that, the performance tends to be saturated when the angular resolution is increased from 7$\times$7 to 9$\times$9. That is because, the the angular information is already sufficient with an angular resolution of 7$\times$7, and a further increase of views can only provide minor performance improvements.

\subsubsection{Comparisons with State-of-the-art Methods}
 We compared our DistgSSR to several state-of-the-art methods, including 4 SISR methods (i.e., VDSR \cite{VDSR}, EDSR \cite{EDSR}, RCAN \cite{RCAN}, ESRGAN \cite{ESRGAN}) and 7 LF image SR methods (i.e., LFBM5D \cite{LFBM5D}, GB \cite{GB}, resLF \cite{resLF}, LFSSR \cite{LFSSR}, LF-ATO \cite{ATO}, LF-InterNet \cite{LF-InterNet}, LF-DFnet \cite{LF-DFnet}). We also include bicubic interpolation as a baseline method. Note that, all deep learning-based SR methods have been retrained on the same training datasets as our method for fair comparison. For simplicity, we only present the results on 5$\times$5 LFs for 2$\times$ and 4$\times$SR.

 \textbf{1) Quantitative results:} Quantitative results are presented in Table \ref{tab:quantitativeSSR}. Our DistgSSR achieves the highest PSNR and SSIM scores on all the 5 datasets for both 2$\times$ and 4$\times$SR. It is worth noting that the improvement of our method is very significant on the HCInew, HCIold and STFgantry datasets for 2$\times$SR (i.e., 0.51dB, 0.47dB, and 0.73dB higher than the second top-performing method on these three datasets, respectively).  That is because, these three datasets were either synthetically rendered \cite{HCInew,HCIold} or captured by a moving camera mounted on a gantry \cite{STFgantry}, and thus have LFs with more complex structures and larger disparity variations than the Lytro datasets \cite{EPFL,INRIA}. By processing features in different subspaces with the disentangling mechanism, our DistgSSR can well handle these complex scenarios while maintaining promising performance for LFs with small baselines.

\textbf{2) Qualitative results:} The qualitative results achieved by different methods for 4$\times$SR are shown in Fig.~\ref{fig:Visual-SSR}. It can be observed from the zoom-in regions that SISR methods cannot reliably recover the missing details. For example, VDSR, EDSR and RCAN produce blurring results with poor details, while ESRGAN hallucinates fake textures with obvious artifacts. In contrast, LF image SR methods achieve significant improvement over SISR methods due to the employment of angular information. Compared with state-of-the-art SISR and LF image SR methods, our DistgSSR can produce images with more faithful details and less artifacts.

\begin{figure*}
\centering
\includegraphics[width=18cm]{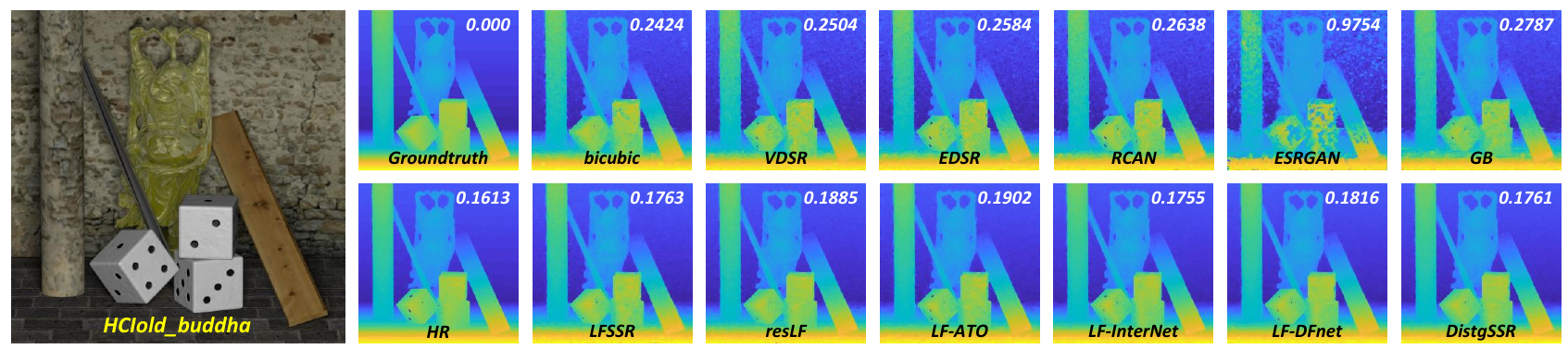}
\caption{Depth estimation results achieved by SPO \cite{SPO} using 4$\times$SR LF images produced by different SR methods. \textcolor{black}{The mean square error multiplied with 100 (i.e., MSE$\times$100) was used as the quantitative metric. Note that, the accuracy is improved by using the LF images produced by our DistgSSR.}}
\label{fig:SSRtoDepth}
\end{figure*}

\textbf{3) Angular consistency:} Since a high-quality SR method should preserve the LF parallax structure and generate angular-consistent LF images, we followed \cite{LFSSR,resLF,ATO} to evaluate the angular consistency of the super-resolved LF images by visualizing their EPI slices. As shown in Fig.~\ref{fig:Visual-SSR}, our method can generate more straight and clear line patterns with less artifacts than other SR methods, which demonstrates that the LF parallax structure is well preserved by our DistgSSR.  Readers are further referred to \href{https://wyqdatabase.s3.us-west-1.amazonaws.com/DistgLF-SpatialSR.mp4}{this video} for a visual comparison of angular consistency. Moreover, since high-resolution and angular-consistent LF images are beneficial to disparity estimation, we further evaluate the angular consistency by using the super-resolved LF images for disparity estimation. We followed \cite{LF-DFnet} to use the 4$\times$SR results and the SPO \cite{SPO} method to estimate the disparity map of the scene \textit{HCIold\_buddha}. It can be observed in Fig.~\ref{fig:SSRtoDepth} that the disparity estimated using the output of our DistgSSR is very close to the results using the original HR LF, which clearly demonstrates the high spatial reconstruction quality and angular consistency achieved by our method.

\begin{table}
\caption{Comparisons of the number of parameters (\#Param.) and FLOPs for 2$\times$ and 4$\times$SR. Note that, FLOPs is calculated on an input LF with an angular resolution of 5$\times$5 and a spatial resolution of 32$\times$32. The PSNR scores are averaged over 5 test datasets.}\label{tab:efficiencySSR}
\centering
\scriptsize
\renewcommand\arraystretch{1.2}
\begin{tabular}{|l|ccc|ccc|}
\hline
\multirow{2}*{Method}&   \multicolumn{3}{c|}{$2\times$}  &   \multicolumn{3}{c|}{$4\times$} \\
\cline{2-7}
  & \tiny{\textbf{\textit{\#Param.}}}  & \tiny{\textbf{\textit{FLOPs}}} & \tiny{\textbf{\textit{PSNR}}} & \tiny{\textbf{\textit{\#Param.}}}  & \tiny{\textbf{\textit{FLOPs}}} & \tiny{\textbf{\textit{PSNR}}}  \\
\hline
EDSR \cite{EDSR}            & 38.6M & 988.9G & 36.04 & 38.9M  & 1017 G & 30.20 \\
RCAN \cite{RCAN}          & 15.3M & 392.8G & 36.11 & 15.4M  & 408.5G & 30.27 \\
LFSSR \cite{LFSSR}        &   \textbf{0.81M} &  25.70G & 37.53 & 1.61M  & 128.4G  & 31.23 \\
resLF \cite{resLF}           &  6.35M & 37.06G & 37.50  & 6.79M & 39.70G  & 31.25 \\
LF-ATO \cite{ATO}      &   1.51M &  597.7G & 38.30 & 1.66M & 687.0G & 31.54 \\
LF-InterNet \cite{LF-InterNet} &   4.80M &  47.46G  & 38.08 & 5.23M & 50.10G  & 31.59 \\
LF-DFnet \cite{LF-DFnet}   & 3.94M &  57.22G  & 38.42 & 3.99M  & 57.31G  & 31.86 \\
\hline
\rowcolor{shadow}
DistgSSR\_32    &  0.88M &  \textbf{16.06G}  & 38.33 & \textbf{0.90M}  & \textbf{16.40G}  & 31.64 \\
\rowcolor{shadow}
DistgSSR\_64    &  3.53M &  64.11G  & \textbf{38.92} & 3.58M  & 65.41G  & \textbf{32.11} \\
\hline
\end{tabular}
\end{table}

\textbf{4) Efficiency:} We compare our DistgSSR to several competitive methods in terms of \textcolor{black}{the number of parameters, FLOPs, and PSNR scores}. As shown in Table~\ref{tab:efficiencySSR}, our DistgSSR achieves the highest PSNR scores with a small number of parameters and reasonable FLOPs. It is worth noting that, even the feature depth of our model is halved to 32, the performance of our method (i.e., DistgSSR\_32) is still better than LF-ATO and LF-InterNet. The competitive performance demonstrates the high efficiency of our DistgSSR.

\begin{figure*}[t]
\centering
\includegraphics[width=16cm]{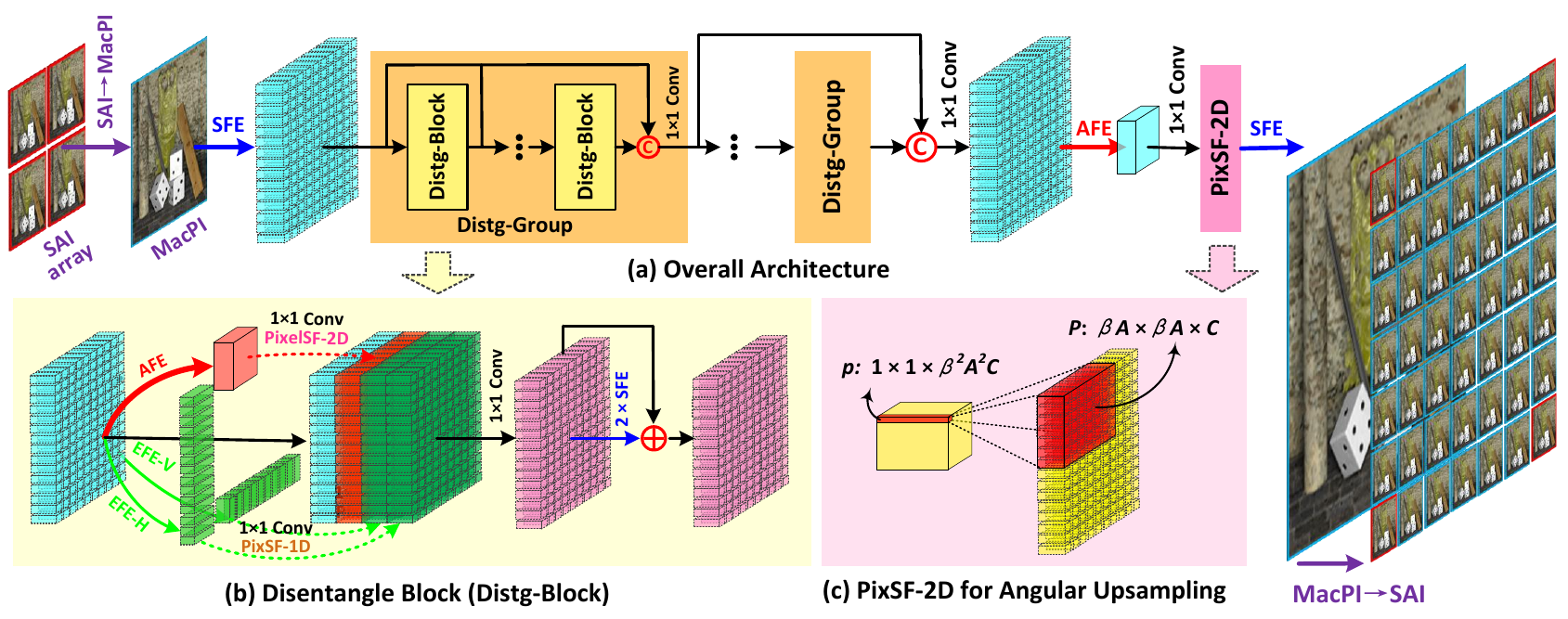}
\caption{{An overview of our DistgASR network. Here, we take the 2$\times$2$\rightarrow$7$\times$7 angular SR as an example to illustrate the network structure.}
\label{DistgASRnet}}
\end{figure*}

\section{DistgASR: Disentangling Mechanism for Angular Super-Resolution}\label{sec:DistgASR}
In this section, we apply our disentangling mechanism to LF angular SR and propose a network named DistgASR. The network design and experimental results are presented in the following subsections.

\subsection{Network Design}
\subsubsection{Overview}
The architecture of  our DistgASR is shown in Fig.~\ref{DistgASRnet}. Our network takes a sparsely-sampled SAI array $\mathcal{I}^{\text{in}}_{\text{SAIs}} \in \mathbb{R}^{AH \times AW}$ as its input to reconstruct a densely-sampled SAI array  $\mathcal{I}^{\text{out}}_{\text{SAIs}} \in \mathbb{R}^{\beta AH \times \beta AW}$, where $\beta$ represents the angular upsampling factor (e.g., $\beta$=$\frac{7}{2}$ for 2$\times$2$\rightarrow$7$\times$7 angular SR). Similar to DistgSSR, our DistgASR also converts the input SAI array into a MacPI $\mathcal{I}^{\text{in}}_{\text{MacPI}} \in \mathbb{R}^{AH \times AW}$ for spatial, angular and EPI feature extraction. Our DistgASR network consists of four disentangling groups (i.e., Distg-Groups), with each Distg-Group comprising of four disentangling blocks (i.e., Distg-Blocks). The output features of each Distg-Block and Distg-Group are respectively concatenated and fed to 1$\times$1 convolutions for multi-stage information fusion.

\subsubsection{Distg-Block for ASR}
Similar to DistgSSR, the Distg-Block is also the basic module of our DistgASR, in which the spatial, angular and EPI information is separately processed and then incorporated. However, the structure of the Distg-Block in our DistgASR is slightly different from that in DistgSSR due to the special characteristics of angular SR. \textbf{First,} the angular information is very important since a large number of novel views need to be reconstructed from a small number of input views. \textbf{Second,} the disparities between two adjacent input views are much larger than those in spatial SR. By considering the aforementioned characteristics, we design our Distg-Block to preserve useful angular and EPI information.

The structure of Distg-Block is shown in Fig.~\ref{DistgASRnet}(b). Given the input MacPI feature $\mathcal{F}_{\text{in}} \in \mathbb{R}^{AH \times AW \times C}$ \textcolor{black}{($C$ is the number of channels and is set to 64 in our DistgASR)} , an angular branch and two EPI branches are first used to disentangle angular and EPI information, respectively. Note that, both angular and EPI branches have the same structure as in DistgSSR but produce output features with a channel number of $C$. After angular and EPI feature extraction, the intermediate features $\mathcal{F}_{\text{Ang}}$, $\mathcal{F}_{\text{Epi-H}}$, $\mathcal{F}_{\text{Epi-V}}$ are concatenated with the input feature $\mathcal{F}_{\text{in}}$ and further fed into a 1$\times$1 convolution for channel reduction. Finally, a spatial residual block with two SFEs is employed for spatial information incorporation.

\subsubsection{Angular Upsampling}
After deep feature disentanglement and incorporation, the MacPI feature needs to be upsampled to the target angular resolution. However, since the upsampling factor $\beta$ in our DistgASR can be non-integer (e.g., $\beta$=$\frac{7}{2}$ for 2$\times$2$\rightarrow$7$\times$7 angular SR), it is infeasible to achieve non-integer upsampling by directly applying pixel-shuffling to the MacPI feature or the SAI array feature. To handle this problem, we introduce a downsample-upsample approach for angular upsampling.

Given the MacPI feature $\mathcal{F}_{\text{fuse}} \in \mathbb{R}^{AH \times AW \times C}$ produced by the 1$\times1$ convolution after the last Distg-Group, an AFE is first employed to generate an angular-downsampled feature $\mathcal{F}_{\text{down}} \in \mathbb{R}^{H \times W \times C}$. Then, a 1$\times$1 convolution is applied to $\mathcal{F}_{\text{down}}$ for channel expansion and produce a feature $\mathcal{F}_{\text{pre-up}} \in \mathbb{R}^{H \times W \times \beta^{2}A^{2}C}$ for angular upsampling. Finally, a 2D pixel-shuffling layer is applied to $\mathcal{F}_{\text{pre-up}}$ to generate $\mathcal{F}^{\text{out}}_{\text{MacPI}} \in \mathbb{R}^{\beta AH \times \beta AW}$. With our downsample-upsample approach, $\beta \times$ upsampling is divided into an $A\times$ downsampling and a $\beta A\times$ upsampling. Since $\beta A$ is the output angular resolution and is an integer, the angular upsampling problem is well handled.

It is worth noting that, after the 2D pixel-shuffling operation, a $1$$\times$$1$$\times$$\beta^{2}A^{2}C$ pixel in $\mathcal{F}_{\text{pre-up}}$ corresponds to a macro-pixel of size $\beta A\times \beta A\times C$ in $\mathcal{F}_{\text{out}}$. Consequently, the final output image $\mathcal{I}^{\text{out}}_{\text{SAIs}}$ can be obtained by first applying an SFE to $\mathcal{F}_{\text{out}}$ for channel reduction, and then performing the MacPI-to-SAI conversion.

\subsection{Experiments}
\subsubsection{Datasets and Implementation Details}
\textcolor{black}{Following \cite{FS-GAF}, we use both synthetic LF datasets (i.e., the HCInew \cite{HCInew} and HCIold \cite{HCIold} datasets) and real-world LF datasets (i.e., the 30scenes \cite{K30} and STFlytro \cite{STFlytro} datasets) for experiments. The split of training and test sets was identical to that in \cite{FS-GAF}. That is, 20 synthetic scenes and 100 real-world scenes were used for training, while 4 scenes from the HCInew dataset \cite{HCInew}, 5 scenes from the HCIold dataset \cite{HCIold}, 30 scenes from the 30scenes dataset \cite{K30}, 25 scenes from the  \textit{Occlusion} category and 15 scenes from the \textit{Reflective} category in the STFlytro dataset \cite{STFlytro} were used for test.} For simplicity, we followed most existing works \cite{P4DCNN,LFASR-geo,FS-GAF,Yeung2018fast} to perform 2$\times$2$\rightarrow$7$\times$7 angular SR. To generate training and test samples, we angularly cropped the central 7$\times$7 SAIs and used their 2$\times$2 corner views as inputs to reconstruct the remaining views. During training phase, we spatially cropped each SAI into patches of 64$\times$64, and generated around 1.5$\times 10^{4}$ training samples for each of the synthetic and real-world datasets. Similar to spatial SR, we performed random horizontal flipping, vertical flipping and 90-degree rotation for data augmentation.

Our network was trained with an L1 loss, and optimized using the Adam method \cite{Adam} with $\beta_1 = 0.9$, $\beta_2 = 0.999$ and a batch size of 4. The initial learning rate was set to 2$\times$10$^{-4}$ and decreased by a factor of 0.5 for every 15 epochs. The training was stopped after 70 epochs. All experiments were conducted on a PC with two Nvidia RTX 2080Ti GPUs.

We used PSNR and SSIM values calculated on the Y channel images for quantitative evaluation. We first calculated the PSNR and SSIM values on the reconstructed views (totally 45 views for 2$\times$2$\rightarrow$7$\times$7 ASR), and then averaged these values to obtain the score of the scene. The score of a dataset is calculated by further averaging the scores of its scenes.

\subsubsection{Ablation Study}
In this subsection, we validate the effectiveness of the disentangling mechanism in angular SR by selectively removing the spatial, angular and EPI branches from our DistgASR. Results are shown in Table~\ref{tab:ablationASR}.

\textbf{Spatial Only}. We introduced \textit{model-1} by using the spatial branch only for angular SR. Without using angular and EPI branches, \textit{model-1} cannot incorporate any angular information and thus suffers a decrease of  3.16dB in average PSNR as compared to DistgASR. Comparative results demonstrate that angular information is very important for angular SR.

\textbf{Angular Only}. We introduced \textit{model-2} by using the angular branch only for angular SR. Consequently, \textit{model-2} cannot incorporate any spatial information and thus suffers a drop of 2.33dB in average PSNR as compared to DistgASR. That is, the spatial information is also of great importance and should be well incorporated for angular SR.

\begin{table}
\caption{Comparative PSNR results achieved by several variants of our DistgASR for 2$\times$2$\rightarrow$7$\times$7 ASR. Note that, the feature depths of different variants were adjusted to make their model size comparable.}\label{tab:ablationASR}
\renewcommand\arraystretch{1.2}
\vspace{-0.2cm}
\centering
\scriptsize
\begin{tabular}{|c|ccc|c|ccc|c|}
\hline
 & \tiny{\textbf{\textit{Spa}}} & \tiny{\textbf{\textit{Ang}}} & \tiny{\textbf{\textit{EPI}}} & \tiny{\textbf{\textit{\#Param.}}} &  \tiny{\textbf{\textit{30scenes}}} &  \tiny{\textbf{\textit{Occlusions}}} &  \tiny{\textbf{\textit{Reflective}}} &  \tiny{\textbf{\textit{Average}}} \\
\hline
 1 & $\checkmark$  &                          &                          &  2.70M&  39.86 & 36.43 & 36.48 & 37.59 \\
 2 &                          & $\checkmark$  &                          & 2.72M & 40.31 & 37.00 & 37.96 & 38.42 \\
 3 & $\checkmark$  & $\checkmark$  &                          & 2.72M& 42.65 & 38.93 & 38.08 & 39.89  \\
 4 & $\checkmark$  & $\checkmark$  & $\checkmark ^{\star}$      & 3.07M    & 43.50 & 39.32 & 38.97 & 40.60 \\
\hline
\end{tabular}

\vspace{0.05cm}
\begin{tabular}{|p{2.4cm}<{\centering}|p{0.66cm}<{\centering}|p{0.73cm}<{\centering}p{0.83cm}<{\centering}p{0.8cm}<{\centering}|p{0.62cm}<{\centering}|}
\hline
 \textit{DistgASR} & \textbf{2.68M}   & \textbf{43.67} & \textbf{39.46} & \textbf{39.11} & \textbf{40.75} \\
\hline
\end{tabular}
\leftline{$^{\star}$Weight sharing is not performed between the two EPI branches in \textit{model-4}.}
\end{table}

\textbf{Spatial+Angular}. We validate the effectiveness of the EPI branches by removing them from our DistgASR. Although both spatial and angular information were used by this variant, \textit{model-3} still suffers a decrease of 0.86dB in average PSNR as compared to DistgASR. That is because, in angular SR, the input views are sparsely sampled and the disparities among adjacent views are several times larger than those in spatial SR. Due to the large disparities, the corresponding pixels of different views may locate in different macro-pixels in MacPI and cannot be effectively convolved by AFEs. With our EFEs, the LFs can be further disentangled into EPI subspaces and the disparity issue can be well handled by performing convolution on EPIs.

\textbf{EPI branch w/o weight sharing}. We introduced \textit{model-4} by removing weight sharing between horizontal and vertical EPI branches. Similar to spatial SR, the performance of \textit{model-4} is slightly inferior to the main model. It further demonstrates that the weight sharing strategy can further regularize our EPI branches to fit the LF parallax structure.

\begin{figure*}
\centering
\includegraphics[width=18cm]{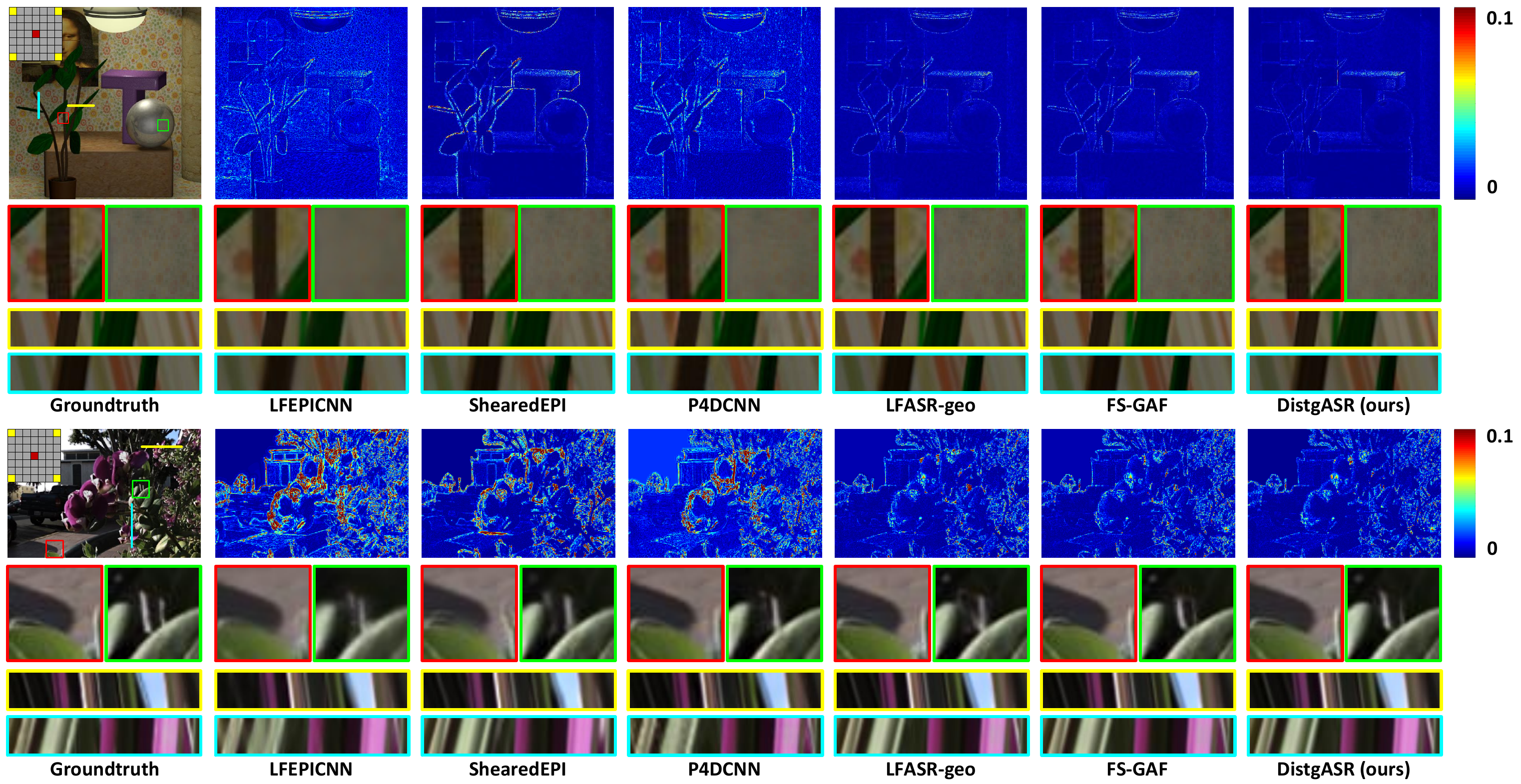}
\caption{Visual results achieved by different methods on scenes \textit{monasRoon} (top) \cite{HCIold} and \textit{IMG\_1555} (bottom) \cite{K30} for 2$\times$2$\rightarrow$7$\times$7 angular SR. Here, we show the error maps of the reconstructed center view images, along with two zoom-in regions and horizontal$/$vertical EPIs.}
\label{fig:Visual-ASR}
\end{figure*}

\subsubsection{Comparisons with State-of-the-art Methods}
We compare our method to 7 state-of-the-art methods including LFEPICNN \cite{LFEPICNN17}, ShearedEPI \cite{ShearedEPI}, P4DCNN \cite{P4DCNN}, Kalantari et al. \cite{K30}, Yeung et al. \cite{Yeung2018fast}, LFASR-geo \cite{LFASR-geo} and FS-GAF \cite{FS-GAF}. All the compared methods except LFEPICNN\footnote{Since the training code of LFEPICNN \cite{LFEPICNN17} is unavailable, we followed \cite{LFASR-geo,FS-GAF} to directly used its released model for inference.} were trained on the same datasets as our method for fair comparison.

\textbf{Quantitative results:}
As shown in Table~\ref{tab:ASR-Quantitative}, the performance of LFEPICNN \cite{LFEPICNN17}, ShearedEPI \cite{ShearedEPI} and P4DCNN \cite{P4DCNN} are inferior to other methods since they only extract features from EPIs. Since the input EPI only contains 2 rows or columns of pixels for 2$\times$2$\rightarrow$7$\times$7 angular SR, it is difficult for these methods to recover the intermediate linear structures without using the spatial context information. In contrast, LFASR-geo \cite{LFASR-geo}, FS-GAF \cite{FS-GAF} and the method proposed by Kalantari et al. \cite{K30} synthesize novel views by performing disparity estimation and feature warping, and achieve much better performance than EPI-based methods \cite{LFEPICNN17,ShearedEPI,P4DCNN}. Among all the compared methods, our DistgASR achieves the highest SSIM scores on all the 5 datasets and the highest PSNR scores on 4 of the 5 datasets (less competitive than FS-GAF on the HCInew dataset). By disentangling input LFs into spatial, angular and EPI subspaces, our DistgASR can fully incorporate the multi-dimensional information and learn the LF structure to accurately reconstruct missing views without explicit disparity estimation.

\begin{table}
\caption{
PSNR and SSIM values achieved by different methods for $2\times2 \rightarrow 7\times7$ angular SR. The best results are in \textbf{bold faces}.}\label{tab:ASR-Quantitative}
\renewcommand\arraystretch{1.2}
\centering
\scriptsize
\begin{tabular}{p{1.27cm}<{\centering}|p{1cm}<{\centering} p{1cm}<{\centering} p{1cm}<{\centering} p{1cm}<{\centering} p{1.05cm}<{\centering}}
\hline
 \tiny{\textbf{\textit{Method}}} &   \tiny{\textbf{\textit{HCInew}}} &  \tiny{\textbf{\textit{HCIold}}} &  \tiny{\textbf{\textit{30scenes}}} &  \tiny{\textbf{\textit{Occlusion} }}  &   \tiny{\textbf{\textit{Reflective}}} \\
\hline
 \tiny{\textbf{\textit{LFEPICNN} \cite{LFEPICNN17}}}  &
26.64/0.744 & 31.43/0.850 & 33.66/0.918 & 32.72/0.924 & 34.76/0.930 \\
 \tiny{\textbf{\textit{ShearedEPI} \cite{ShearedEPI}}} &
31.84/0.898 & 37.61/0.942 & 39.17/0.975 & 34.41/0.955 & 36.38/0.944 \\
 \tiny{\textbf{\textit{P4DCNN} \cite{P4DCNN}}} &
29.61/0.819 & 35.73/0.898 & 38.22/0.970 & 35.42/0.962 & 35.96/0.942 \\
 \tiny{\textbf{\textit{Kanlantari} \cite{K30}}} &
32.85/0.909 & 38.58/0.944 & 41.40/0.982 & 37.25/0.972 & 38.09/0.953 \\
 \tiny{\textbf{\textit{Yeung} \cite{Yeung2018fast}}} &
32.30/0.900 & 39.69/0.941 & {42.77}/{0.986} & {38.88/0.980} & 38.33/{0.960} \\
 \tiny{\textbf{\textit{LFASR-geo} \cite{LFASR-geo}}} &
34.60/0.937 & 40.84/0.960 & 42.53/0.985 & 38.36/0.977 & 38.20/0.955 \\
 \tiny{\textbf{\textit{FS-GAF} \cite{FS-GAF}}} &
\textbf{37.14}/{0.966} & {41.80/0.974} & 42.75/{0.986} & 38.51/0.979 & {38.35}/0.957 \\
 \tiny{\textbf{\textit{DistgASR}}} &
{34.70}/\textbf{0.974} & \textbf{42.18}/\textbf{0.978} & \textbf{43.67}/\textbf{0.995} & \textbf{39.46}/\textbf{0.991} & \textbf{39.11}/\textbf{0.978} \\
\hline
\end{tabular}
\leftline{Note: All compared methods except LFEPICNN \cite{LFEPICNN17} were retrained on the same}\\
\leftline{datasets as our method.}
\end{table}

\textbf{Qualitative results:}
Figure~\ref{fig:Visual-ASR} shows the qualitative results achieved by different methods. It can be observed from the error maps that the views reconstructed by our method are more close to the groundtruth, and the delicate structures (e.g., edges of leaves in scene \textit{monasRoom}) are well maintained. As shown in the zoom-in regions, during synthesizing novel views, our method can well maintain the detail textures from input views to the target views, while the compared methods suffer from different degrees of blurs or artifacts.

\begin{figure}
\centering
\includegraphics[width=8.8cm]{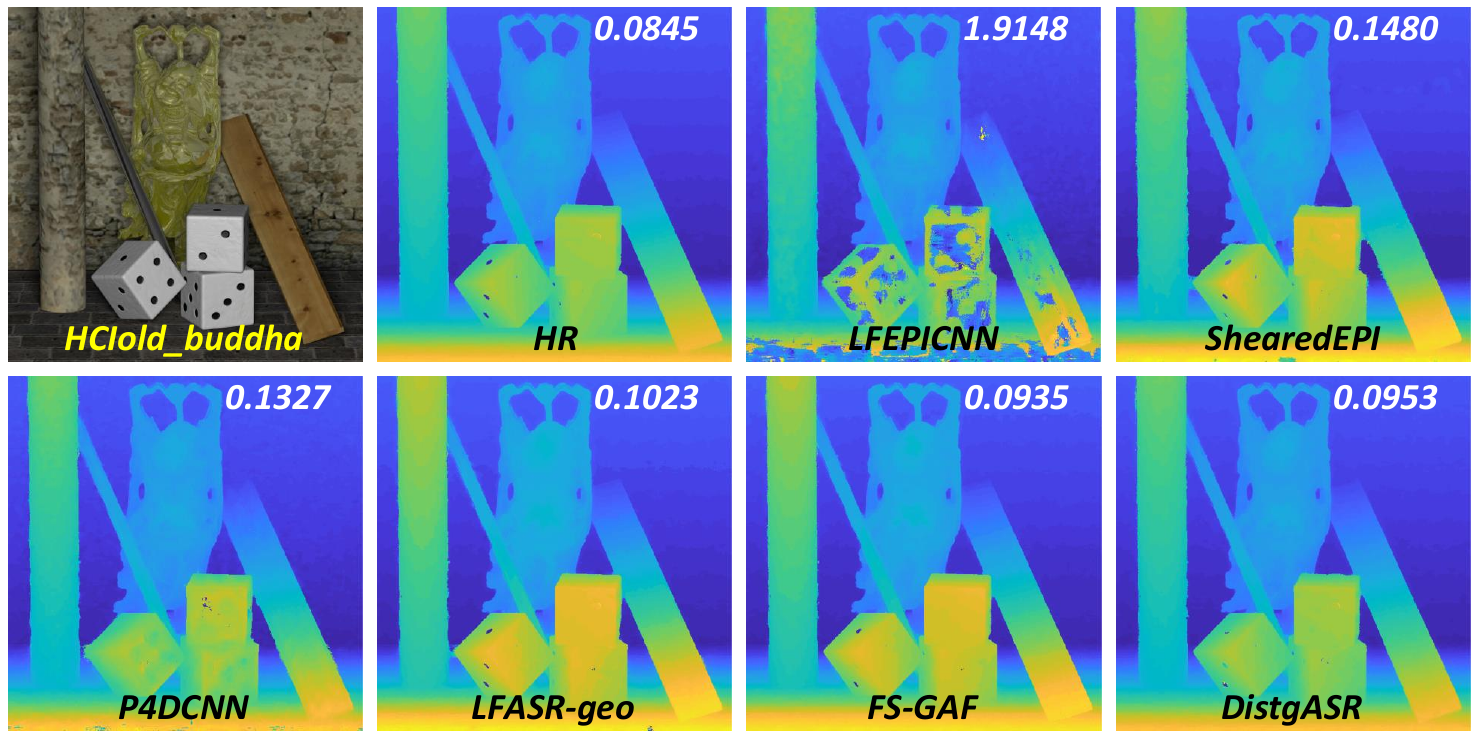}
\caption{\textcolor{black}{Disparity estimation results achieved by SPO \cite{SPO} using LF images produced by different angular SR methods. The mean square error multiplied with 100 (i.e., MSE$\times$100) was used as the quantitative metric. The accuracy is improved by using the LF images produced by our DistgASR.}}
\label{fig:ASRtoDepth}
\end{figure}

\textcolor{black}{\textbf{Angular consistency:}
To evaluate the angular consistency achieved by different methods, we first visualize both horizontal and vertical EPIs of the reconstructed LFs. It can be observed in Fig.~\ref{fig:Visual-ASR} that LFASR-geo, FS-GAF and our method can well preserve the linear parallax structure and produce EPIs close to the groundtruth ones. Then, we used the reconstructed LFs and the SPO method \cite{SPO} for disparity estimation. As shown in Fig.~\ref{fig:ASRtoDepth}, the disparity estimated using the LF reconstructed by our DistgASR is more accurate. It demonstrates the high angular consistency achieved by our method. Finally, we provide a \href{https://wyqdatabase.s3.us-west-1.amazonaws.com/DistgLF-AngularSR.mp4}{demo video} for a visual comparison of angular consistency}.

\section{DistgDisp: Disentangling Mechanism for Disparity Estimation}\label{sec:DistgDisp}
In this section, we apply our disentangling mechanism to LF disparity estimation and propose a network named DistgDisp. The network design and experimental results are introduced in the following subsections.

\subsection{Network Design}
An overview of our DistgDisp is shown in Fig.~\ref{DistgDisp}. Our network takes a MacPI (with an angular resolution of 9$\times$9) as its input and sequentially performs feature extraction, cost volume construction, cost aggregation, and disparity regression. \textcolor{black}{Here, we apply our disentangling mechanism to disparity estimation by using SFE for spatial feature extraction (similar to those in DistgSSR and DistgASR) and extending AFE to disparity-selective angular feature extractors (DS-AFEs) for cost volume construction. The constructed 4D cost volume is further aggregated by eight cascaded 3D convolutions to achieve final disparity regression.}

\subsubsection{Spatial Res-Block for Feature Extraction}
\begin{figure*}[t]
\centering
\includegraphics[width=16cm]{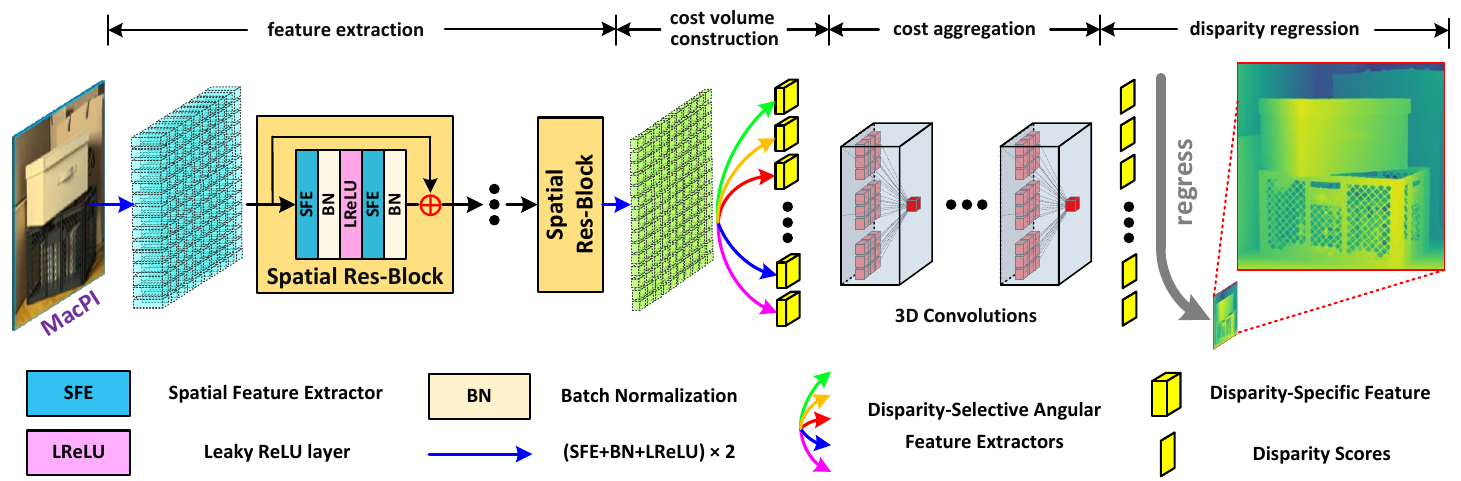}
\caption{{An overview of our DistgDisp. The input 9$\times$9 SAIs are first re-organized into a MacPI and fed to 8 spatial residual blocks for spatial information incorporation. Then, a series of disparity-selective angular feature extractors are introduced to disentangle the disparity information from the MacPI features to generate cost volumes. The generated cost volumes are further aggregated via a 3D hourglass module to regress the final disparity.}
\label{DistgDisp}}
\end{figure*}

Spatial context information is important to disparity estimation, especially for specular and textureless regions \cite{LFAttNet}. Consequently, we design a spatial residual block (i.e., Spatial Res-Block) \textcolor{black}{with batch normalization (BN)} to model the relationship between each pixel and its spatially adjacent pixels. As shown in Fig.~\ref{DistgDisp}, we build our Spatial Res-Block in an ``SFE-BN-LeakyReLU-SFE-BN'' structure, and add the original input feature to the end of the block for local residual learning.

 \subsubsection{DS-AFEs for Cost Volume Construction}
 Cost volume is a widely-used technique in stereo matching to extract long-range correspondence along the epipolar line. In the area of LF disparity estimation, existing methods generally use the ``shift-and-concat'' approach \cite{LFAttNet} to build cost volumes. However, shifting a large number of LF images with a series of pre-defined offset values is computationally inefficient and difficult for parallel processing\footnote{For example, in LFAttNet \cite{LFAttNet}, totally 80 views are shifted by 8 disparity levels, resulting in 640 sequential shifting operations.}. In this paper, we apply our disentangling mechanism to LF disparity estimation and propose disparity-selective angular feature extractors (DS-AFEs) to convolve pixels with specific disparity values for cost volume construction.

\begin{figure}
\centering
\includegraphics[width=8.8cm]{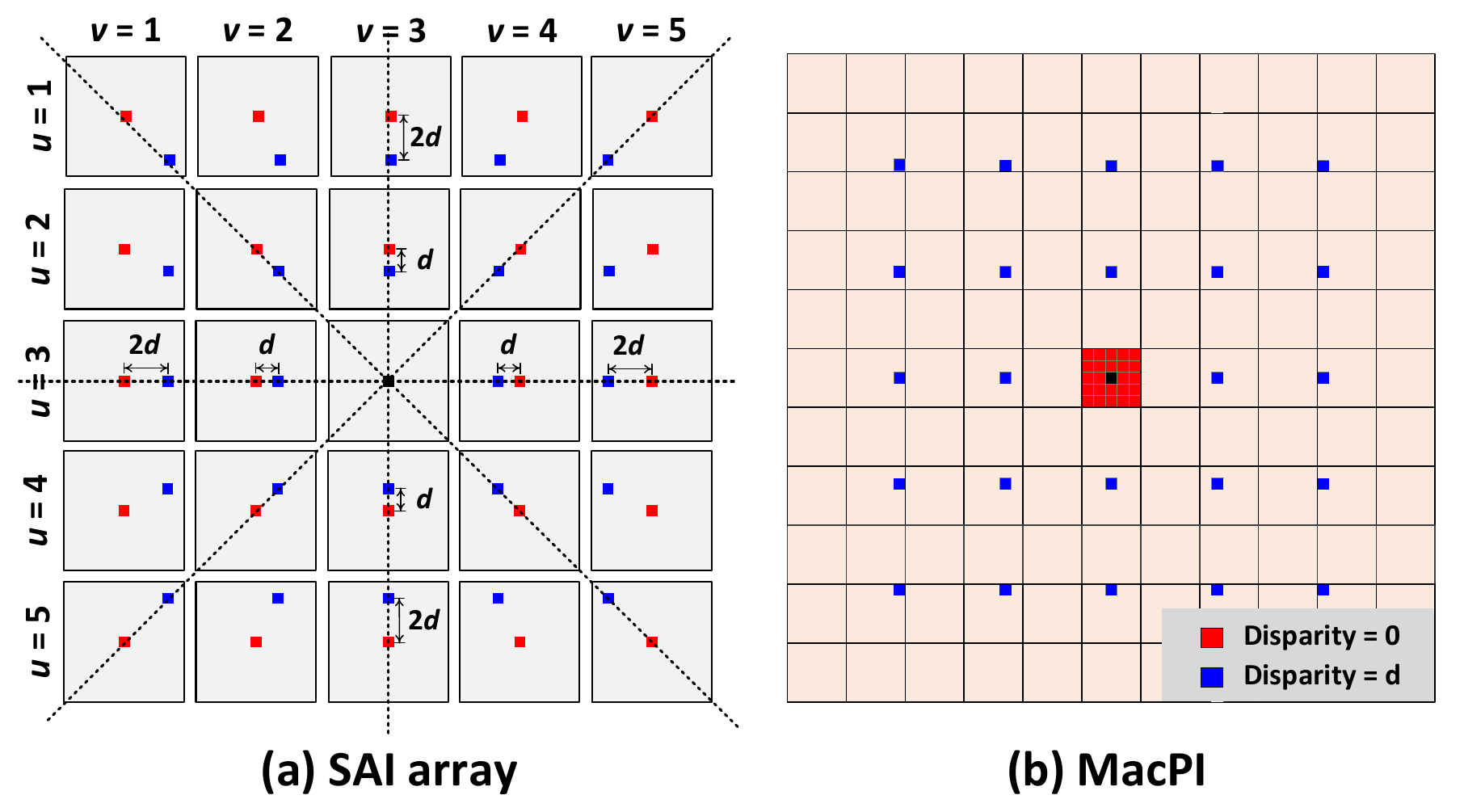}
\caption{{An illustration of our disparity-selective angular feature extractors (DS-AFEs). Here, an LF of size $U$=$V$=5 (i.e., $A$=5), $H$=$W$=11 is used as a toy example and organized both in an SAI array (sub-figure (a)) and a MacPI (sub-figure (b)). Here, the center pixel of the center view is paint in black, and its corresponding pixels in side views are paint in red for disparity=0 and blue for disparity=$d$. When converting the SAI array into MacPI, the pixels under a specific disparity value can be exactly convolved by our DS-AFE with specifically designed dilation and padding values.}
\label{fig:DS-AFE}}
\end{figure}

  Before introducing our DS-AFE, we first illustrate the LF structure using a toy example. As shown in Fig.~\ref{fig:DS-AFE}, an LF of size $U$=$V$=5 (i.e., $A$=5), $H$=$W$=11 can be organized in an SAI array (Fig.~\ref{fig:DS-AFE}(a)) or a MacPI (Fig.~\ref{fig:DS-AFE}(b)). The center pixel of the center view (colored in black) is set as the anchor pixel. As shown in Fig.~\ref{fig:DS-AFE}(a), its corresponding pixels  locate at the center of each side view (colored in red) if the disparity $d$ is 0. Otherwise, its corresponding pixels (colored in blue) have view-dependent offsets from the center of each side view. The offset is calculated as:
     \begin{equation}\label{Eq:offset}
     \left(\Delta h, \Delta w\right)_{u,v} = \left(d\cdot(u_c - u), d\cdot(v_c - v)\right),
     \end{equation}
  where $\left(\Delta h, \Delta w\right)_{u,v}$ represent the vertical and horizontal offsets of view $(u, v)$, $(u_c, v_c)$ represent the angular coordinates of the center view, and $d$ represents the disparity between two adjacent views. When converting the SAI array into a MacPI, these corresponding pixels can be evenly distributed in a square region centered at the anchor pixel (see Fig.~\ref{fig:DS-AFE}(b)), with different dilation rates for different disparity values. Based on this observation, we design our DS-AFE as a series of convolutions performed on MacPIs, with a kernel size of $A$$\times$$A$, a stride of $A$, and specifically designed dilation rates and padding values. For a given preset disparity value $d\in \mathbb{Z}$, the dilation rate $dila$ and padding value $pad$ can be calculated as
    \begin{equation}\label{Eq:dilation}
    \begin{split}
    dila = \left \{
               \begin{array}{ll}
                         d\cdot A - 1,          & d>0, \\
                         -d\cdot A + 1,       & d\leq 0, \\
               \end{array}\right.
    \end{split}
    \end{equation}

    \begin{equation}\label{Eq:padding}
    \begin{split}
    pad = \left \{
               \begin{array}{ll}
                         d\cdot A(A-1)/2 - A + 1,  & d>0, \\
                         -d\cdot A(A-1)/2,             & d\leq 0. \\
               \end{array}\right.
    \end{split}
    \end{equation}

  \textcolor{black}{Based on Eqs.~\ref{Eq:dilation} and \ref{Eq:padding}, we calculate the dilation rates and padding values for each disparity candidate}. Note that, if the preset disparity value is 0, our DS-AFE  degeneralizes to an AFE (which convolves pixels in a macro-pixel). With our DS-AFEs, pixels from each view under a specific disparity can be exactly convolved. Consequently, the cost volumes can be easily constructed by performing DS-AFEs with a series of predefined disparity values.  In this paper, we followed \cite{LFAttNet} to set 9 disparity levels (ranging from -4 to 4). After generating the cost volumes, we concatenate these features into a 5D tensor of size $B\times 9 \times C \times H \times W$, where $B$ and $C$ represent the batch size and the number of channels, respectively.

\begin{table*}
\caption{
\textcolor{black}{Comparative results achieved by different LF disparity estimation methods on the HCI benchmark. The best results are in \textbf{bold faces}.}}\label{tab:QuantitativeDisp}
\centering
\scriptsize
\renewcommand\arraystretch{1.05}
\begin{tabular}{|l|
p{0.5cm}<{\centering}p{0.52cm}<{\centering}p{0.52cm}<{\centering}p{0.52cm}<{\centering}| p{0.5cm}<{\centering}p{0.52cm}<{\centering}p{0.52cm}<{\centering}p{0.52cm}<{\centering}| p{0.5cm}<{\centering}p{0.52cm}<{\centering}p{0.52cm}<{\centering}p{0.52cm}<{\centering}|
p{0.5cm}<{\centering}p{0.52cm}<{\centering}p{0.52cm}<{\centering}p{0.52cm}<{\centering}|}
\hline
\hline
\multirow{2}*{} &   \multicolumn{4}{c|}{\tiny{\textbf{\textit{Backgammon}}}} & \multicolumn{4}{c|}{\tiny{\textbf{\textit{Dots}}}} & \multicolumn{4}{c|}{\tiny{\textbf{\textit{Pyramids}}}} & \multicolumn{4}{c|}{\tiny{\textbf{\textit{Stripes}}}} \\
\cline{2-17}
& \tiny{\textbf{\textit{BP07}}} &  \tiny{\textbf{\textit{BP03}}} &  \tiny{\textbf{\textit{BP01}}} & \tiny{\textbf{\textit{MSE}}} &
 \tiny{\textbf{\textit{BP07}}} &  \tiny{\textbf{\textit{BP03}}} &  \tiny{\textbf{\textit{BP01}}} & \tiny{\textbf{\textit{MSE}}} &
 \tiny{\textbf{\textit{BP07}}} &  \tiny{\textbf{\textit{BP03}}} &  \tiny{\textbf{\textit{BP01}}} & \tiny{\textbf{\textit{MSE}}} &
 \tiny{\textbf{\textit{BP07}}} &  \tiny{\textbf{\textit{BP03}}} &  \tiny{\textbf{\textit{BP01}}} & \tiny{\textbf{\textit{MSE}}}  \\
\hline
\textit{CAE} \cite{CAE} &
3.924 & \textbf{4.313} & {17.32} & 6.074 &
12.40 & 42.50 & 83.70 & 5.082 &
1.681 & 7.162 & 27.54 & 0.048 &
7.872 & 16.90 & 39.95 & 3.556 \\

\textit{PS-RF} \cite{PS-RF} &
7.142 & 13.94 & 74.66 & 6.892 &
7.975 & 17.54 & 78.80 & 8.338 &
\textbf{0.107} & 6.235 & 83.23 & 0.043 &
{2.964} & {5.790} & 41.65 & 1.382 \\

\textit{SPO} \cite{SPO} &
3.781 & 8.639 & 49.94 & 4.587 &
16.27 & 35.06 & 58.07 & 5.238 &
0.861 & 6.263 & 79.20 & 0.043 &
14.97 & 15.46 & 21.87 & 6.955 \\


\textit{OBER-crossANP} \cite{OBER} &
{3.413} & {4.952} & \textbf{13.66} & 4.700 &
\textbf{0.974} & 37.66 & 73.13 & 1.757 &
0.364 & 1.130 & {8.171} & {0.008} &
3.065 & 9.352 & 44.72 & 1.435 \\

\textit{EPN+OS+GC} \cite{EPN} &
\textbf{3.328} & 10.56 & 55.98 & 3.699 &
39.25 & 82.74 & 84.91 & 22.37 &
0.242 & 3.169 & 28.56 & 0.018 &
18.54 & 19.60 & 28.17 & 8.731  \\

\textit{Epinet-fcn} \cite{EPINET} &
3.580 & 6.289 & 20.89 & {3.629} &
3.183 & {12.73} & {41.05} & {1.635} &
0.192 & {0.913} & 11.87 & {0.008} &
\textbf{2.462} & \textbf{3.115} & \textbf{15.67} & {0.950} \\

\textit{EPI-Shift} \cite{EPI-Shift} &
22.89 & 40.53 & 70.58 & 12.79 &
43.92 & 53.18 & 74.55 & 13.15 &
1.242 & 7.315 & 40.48 & 0.037 &
22.72 & 47.70 & 78.95 & 1.686 \\

\textit{EPI\_ORM} \cite{ORM} &
3.988 & 7.238 & 34.32 & \textbf{3.411} &
36.10 & 47.93 & 65.71 & 14.48 &
0.324 & 1.301 & 19.06 & 0.016 &
6.871 & 13.94 & 55.14 & 1.744 \\


\textit{DistgDisp} (ours) &
5.824 & 10.54 & 26.17 & 4.712 &
{1.826} & \textbf{4.464} & \textbf{25.37} & \textbf{1.367} &
{0.108} & \textbf{0.539} & \textbf{4.953} & \textbf{0.004} &
3.913 & 6.885 & {19.25} & \textbf{0.917} \\
\hline
\hline

\multirow{2}*{} &   \multicolumn{4}{c|}{\tiny{\textbf{\textit{Boxes}}}} & \multicolumn{4}{c|}{\tiny{\textbf{\textit{Cotton}}}} & \multicolumn{4}{c|}{\tiny{\textbf{\textit{Dino}}}} & \multicolumn{4}{c|}{\tiny{\textbf{\textit{Sideboard}}}} \\
\cline{2-17}
& \tiny{\textbf{\textit{BP07}}} &  \tiny{\textbf{\textit{BP03}}} &  \tiny{\textbf{\textit{BP01}}} & \tiny{\textbf{\textit{MSE}}} &
 \tiny{\textbf{\textit{BP07}}} &  \tiny{\textbf{\textit{BP03}}} &  \tiny{\textbf{\textit{BP01}}} & \tiny{\textbf{\textit{MSE}}} &
 \tiny{\textbf{\textit{BP07}}} &  \tiny{\textbf{\textit{BP03}}} &  \tiny{\textbf{\textit{BP01}}} & \tiny{\textbf{\textit{MSE}}} &
 \tiny{\textbf{\textit{BP07}}} &  \tiny{\textbf{\textit{BP03}}} &  \tiny{\textbf{\textit{BP01}}} & \tiny{\textbf{\textit{MSE}}}  \\
\hline
\textit{CAE} \cite{CAE} &
17.89 & 40.40 & 72.69 & 8.424 &
3.369 & 15.50 & 59.22 & 1.506 &
4.968 & 21.30 & 61.06 & 0.382 &
9.845 & 26.85 & 56.92 & 0.876  \\

\textit{PS\_RF} \cite{PS-RF} &
18.95 & 35.23 & 76.39 & 9.043 &
2.425 & 14.98 & 70.41 & 1.161 &
4.379 & 16.44 & 75.97 & 0.751 &
11.75 & 36.28 & 79.98 & 1.945 \\

\textit{SPO} \cite{SPO} &
15.89 & 29.52 & 73.23 & 9.107 &
2.594 & 13.71 & 69.05 & 1.313 &
2.184 & 16.36 & 69.87 & 0.310 &
9.297 & 28.81 & 73.36 & 1.024  \\


\textit{OBER-crossANP} \cite{OBER} &
\textbf{10.76} & \textbf{17.92} & {44.96} & 4.750 &
1.108 & 7.722 & 36.79 & 0.555 &
2.070 & 6.161 & 22.76 & 0.336 &
5.671 & 12.48 & {32.79} & 0.941  \\

\textit{EPN+OS+GC} \cite{EPN} &
15.30 & 29.01 & 67.35 & 9.314 &
2.060 & 9.767 & 54.85 & 1.406 &
2.877 & 12.79 & 58.79 & 0.565 &
7.997 & 23.87 & 66.35 & 1.744  \\

\textit{Epinet-fcn} \cite{EPINET} &
{12.84} & {19.76} & 49.04 & 6.240 &
{0.508} & {2.310} & {28.06} & {0.191} &
\textbf{1.286} & \textbf{3.452} & {22.40} & {0.167} &
{4.801} & {12.08} & 41.88 & 0.827 \\


\textit{EPI-Shift} \cite{EPI-Shift} &
25.95 & 44.14 & 74.36 & 9.790 &
2.176 & 10.68 & 46.86 & 0.475 &
5.964 & 22.14 & 64.16 & 0.392 &
11.80 & 36.64 & 73.42 & 1.261  \\

\textit{EPI\_ORM} \cite{ORM} &
13.37 & 25.33 & 59.68 & {4.189} &
0.856 & 5.564 & 42.94 & 0.287 &
2.814 & 8.993 & 41.04 & 0.336 &
5.583 & 14.61 & 52.59 & {0.778} \\

\textit{DistgDisp} (ours) &
13.31 & 21.13 & \textbf{41.62} & \textbf{3.325} &
\textbf{0.489} & \textbf{1.478} & \textbf{7.594} & \textbf{0.184} &
{1.414} & {4.018} & \textbf{20.46} & \textbf{0.099} &
\textbf{4.051} & \textbf{9.575} & \textbf{28.28} & \textbf{0.713} \\
\hline
\hline
\multirow{2}*{} &   \multicolumn{4}{c|}{\tiny{\textbf{\textit{Bedroom}}}} & \multicolumn{4}{c|}{\tiny{\textbf{\textit{Bicycle}}}} & \multicolumn{4}{c|}{\tiny{\textbf{\textit{Herbs}}}} & \multicolumn{4}{c|}{\tiny{\textbf{\textit{Origami}}}} \\
\cline{2-17}
& \tiny{\textbf{\textit{BP07}}} &  \tiny{\textbf{\textit{BP03}}} &  \tiny{\textbf{\textit{BP01}}} & \tiny{\textbf{\textit{MSE}}} &
 \tiny{\textbf{\textit{BP07}}} &  \tiny{\textbf{\textit{BP03}}} &  \tiny{\textbf{\textit{BP01}}} & \tiny{\textbf{\textit{MSE}}} &
 \tiny{\textbf{\textit{BP07}}} &  \tiny{\textbf{\textit{BP03}}} &  \tiny{\textbf{\textit{BP01}}} & \tiny{\textbf{\textit{MSE}}} &
 \tiny{\textbf{\textit{BP07}}} &  \tiny{\textbf{\textit{BP03}}} &  \tiny{\textbf{\textit{BP01}}} & \tiny{\textbf{\textit{MSE}}}  \\
\hline
\textit{CAE} \cite{CAE} &
5.788 & 25.36 & 68.59 & 0.234 &
11.22 & 23.62 & 59.64 & 5.135 &
9.550 & 23.16 & 59.24 & 11.67 &
10.03 & 28.35 & 64.16 & 1.778 \\

\textit{PS\_RF} \cite{PS-RF} &
6.015 & 22.45 & 80.68 & 0.288 &
17.17 & 32.32 & 79.80 & 7.926 &
10.48 & 21.90 & 66.47 & 15.25 &
13.57 & 36.45 & 80.32 & 2.393 \\

\textit{SPO} \cite{SPO} &
4.864 & 23.53 & 72.37 & 0.209 &
10.91 & 26.90 & 71.13 & 5.570 &
8.260 & 30.62 & 86.62 & 11.23 &
11.69 & 32.71 & 75.58 & 2.032 \\


\textit{OBER-crossANP} \cite{OBER} &
3.329 & 9.558 & {28.91} & {0.185} &
\textbf{8.683} & \textbf{16.17} & {37.83} & 4.314 &
{7.120} & {14.06} & {36.83} & 10.44 &
8.665 & 20.03 & {42.16} & 1.493  \\

\textit{EPN+OS+GC} \cite{EPN} &
7.543 & 16.76 & 58.93 & 1.188 &
11.60 & 24.86 & 64.10 & 6.411 &
9.190 & 25.72 & 67.13 & 11.58 &
10.75 & 27.09 & 67.35 & 10.09  \\

\textit{Epinet-fcn} \cite{EPINET} &
{2.403} & {6.921} & 33.99 & 0.213 &
9.896 & 18.05 & 46.37 & 4.682 &
12.10 & 28.95 & 62.67 & 9.700 &
 {5.918} & {14.37} & 45.93 & {1.466} \\


\textit{EPI-Shift} \cite{EPI-Shift} &
8.297 & 21.51 & 55.45 & 0.284 &
20.79 & 39.59 & 68.48 & 6.920 &
14.19 & 26.66 & 56.98 & 17.01 &
11.52 & 33.75 & 73.45 & 1.690  \\

\textit{EPI\_ORM} \cite{ORM} &
5.492 & 14.66 & 51.02 & 0.298 &
11.12 & 21.20 & 51.22 & {3.489} &
8.515 & 24.60 & 68.79 & \textbf{4.468} &
8.661 & 22.95 & 56.57 & 1.826 \\

\textit{DistgDisp} (ours) &
\textbf{2.349} & \textbf{5.925} & \textbf{17.66} & \textbf{0.111} &
{9.856} & {17.58} & \textbf{35.72} & \textbf{3.419} &
\textbf{6.846} & \textbf{12.44} & \textbf{24.44} & {6.846} &
\textbf{4.270} & \textbf{9.816} & \textbf{28.42} & \textbf{1.053}  \\
\hline
\end{tabular}
\end{table*}

\subsubsection{Cost Aggregation and Disparity Regression}
Given the matching costs generated by our DS-AFEs, we followed \cite{LFAttNet} to cascade eight 3D convolutions with a kernel size of 3$\times$3$\times$3 and a stride of 1 for cost aggregation. Finally, a 3D tensor $\mathcal{F}_\text{final}\in \mathbb{R}^{D\times H\times W}$ is generated, and the disparity is regressed according to
\begin{equation}\label{Eq:regress}
  \hat{d} =  \sum_{d\in D} d \times \text{softmax}(\mathcal{F}_\text{final}),
\end{equation}
where $\hat{d}$ denotes the final estimated disparity, $D=\{-4,-3,\cdots,3,4\}$ is the set of the preset disparity levels. Softmax normalization is performed along the disparity axis of  $\mathcal{F}_\text{final}$.

\subsection{Experiments}
In this section, we first introduce the datasets and implementation details, then compare our DistgDisp to state-of-the-art LF disparity estimation methods.

\subsubsection{Datasets and Implementation Details}
We used the 4D LF benchmark \cite{HCInew} to validate the effectiveness of our method. All LFs in this benchmark have an angular resolution of 9$\times$9 and a spatial resolution of 512$\times$512. We followed \cite{EPINET} to use 16 scenes in the ``Additional'' category for training, 8 scenes in the ``Stratified'' and ``Training'' categories for validation, and 4 scenes in the ``Test'' category for test.

During the training phase, we randomly cropped SAIs into patches of size 64$\times$64, and converted them into gray-scale images. We performed a large variety of data augmentation, including random flipping and rotation, color channel re-distribution, brightness and contrast adjustment, refocusing and  downsampling. Our network was trained using an L1 loss and optimized using the Adam method \cite{Adam} with $\beta_1$=0.9, $\beta_2$=0.999 and a batch size of 12. The initial learning rate was set to 1$\times$10$^{-3}$ and was decreased by a factor of 0.5 for every 3000 epochs. The training was stopped after 15000 epochs. Our model was implemented in PyTorch and trained on a PC with an Nvidia RTX 3090 GPU.

We used the mean square error (MSE)\footnote{The MSE scores reported in this section are multiplied with 100 for better comparison.} and bad pixel ratio (BadPix($\epsilon$)) as quantitative metrics for performance evaluation. BadPix($\epsilon$) measures the percentage of incorrectly estimated pixels whose absolute errors exceeding a predefined threshold ($\epsilon=$0.07, 0.03, 0.01).

\begin{figure}
\centering
\includegraphics[width=8.8cm]{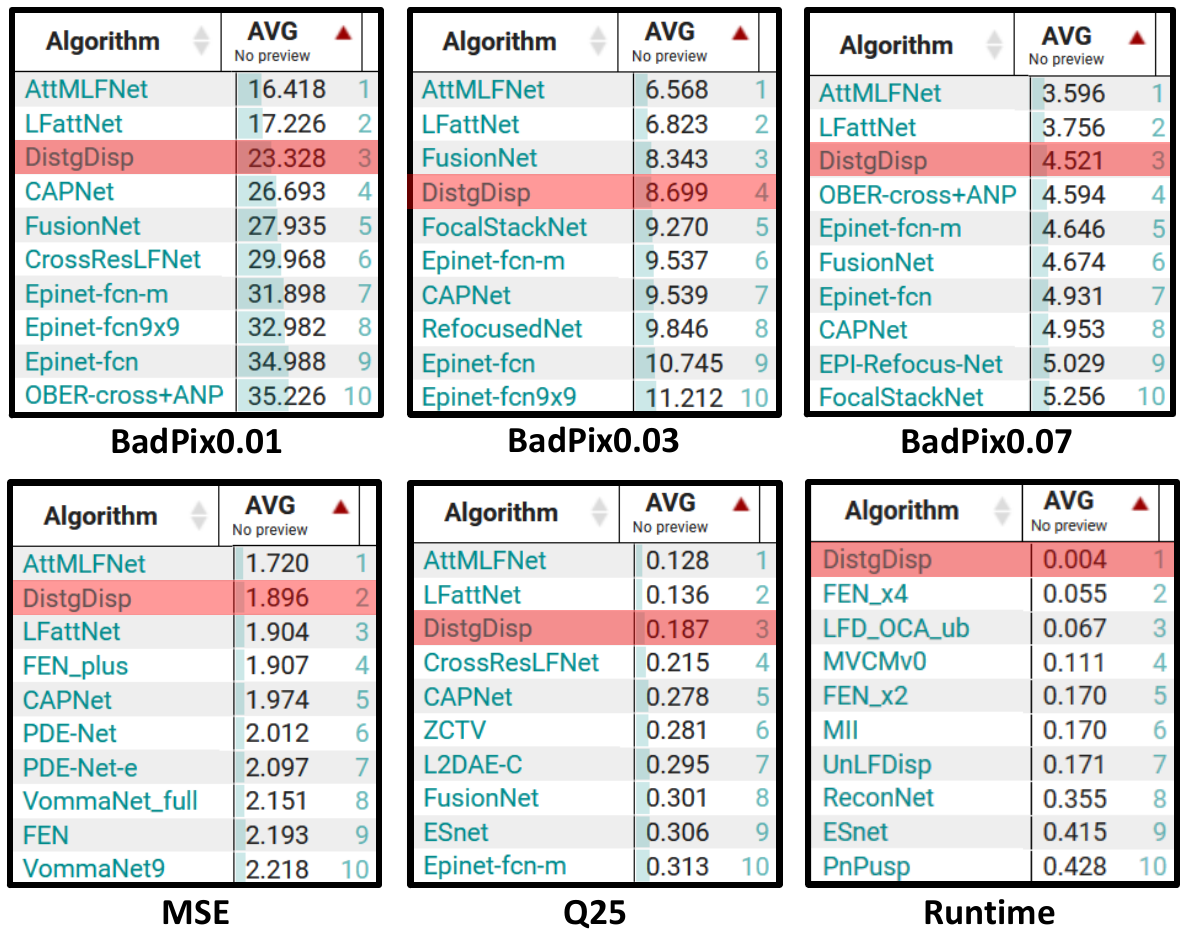}
\caption{{The screenshot of the rankings on the 4D LF benchmark (captured in July 2021). Our \textit{DistgDisp} ranks the second to the fourth place among all the 81 submissions in terms of the average values of five mainly used metrics, and ranks the first place in terms of the average running time.}
\label{fig:Screenshot-Disp}}
\end{figure}

\begin{figure*}[t]
\centering
\includegraphics[width=18cm]{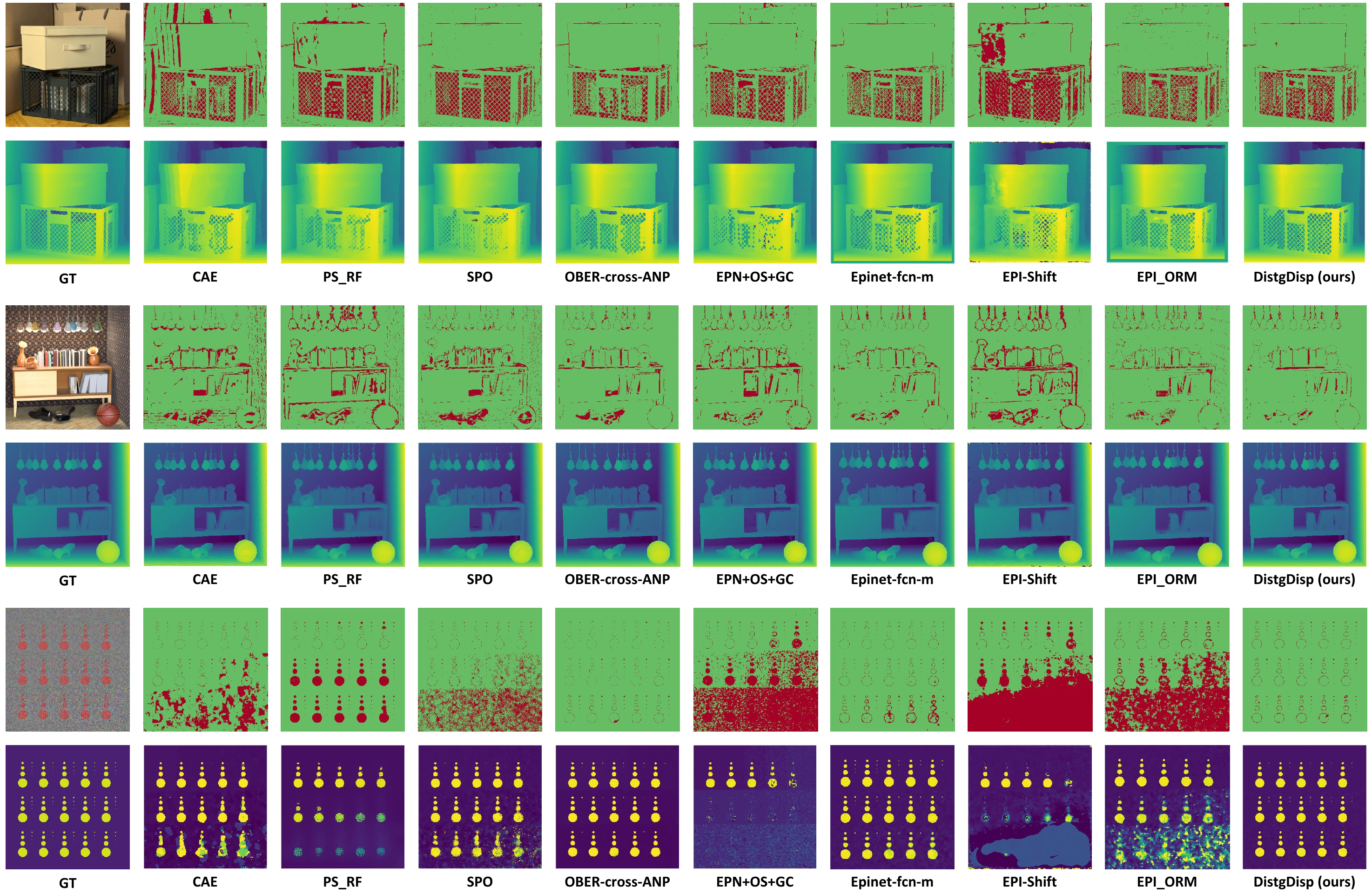}
\caption{{Visual comparisons among different LF disparity estimation methods on the 4D LF benchmark \cite{HCInew}. For each scene, the bottom row shows the estimated disparity maps and the top row shows the corresponding BadPix0.07 maps (pixels with absolute error larger than 0.07 are marked in red).}
\label{fig:Visual-Disp}}
\end{figure*}

\subsubsection{Comparisons with State-of-the-art Methods}
We compare our method to eight state-of-the-art methods, including four traditional methods (i.e., CAE \cite{CAE}, PS\_RF \cite{PS-RF}, SPO \cite{SPO}, OBER-Cross-ANP \cite{OBER}) and four deep learning-based methods (i.e., EPN+OS+GC \cite{EPN}, Epinet-fcn \cite{EPINET}, EPI-Shift \cite{EPI-Shift}, EPI\_ORM \cite{ORM}).

\textbf{1) Quantitative Results:} Table \ref{tab:QuantitativeDisp} shows the quantitative results achieved by different methods on each validation and test scene. Among the 12 scenes, our method obtains the first place on 10 scenes in terms of \textit{MSE} and \textit{BadPix0.01}. To achieve further comparison, we submitted our results to the benchmark and compared our method to all the submissions. Figure~\ref{fig:Screenshot-Disp} shows the top-10 lowest errors$/$running time averaged on the whole validation and test sets. Among all the 81 submissions, our DistgDisp ranks the top four in terms of five major error metrics (i.e., \textit{BadPix0.07}, \textit{BadPix0.03}, \textit{BadPix0.01}, \textit{MSE},  \textit{Q25}). Moreover, our method ranks the first place in terms of running time, i.e., our method is faster than other methods by an order of magnitude. It is worth noting that, the top-ranking methods LFattNet \cite{LFAttNet} and AttMLFNet \cite{AttMLFNet} achieve  accurate disparity estimation using delicately designed feature extraction and attention-based view selection modules. In contrast, our DistgDisp provides a relatively simple baseline using only plain convolutions and residual blocks to achieve comparable performance without bells and whistles. This further demonstrate the effectiveness of our disentangling mechanism in LF disparity estimation.

\textbf{2) Visual Comparison:} Figure~\ref{fig:Visual-Disp} shows some examples of estimated disparity maps and corresponding \textit{BadPix0.07} error maps. It can be observed that the disparities estimated by our method are more close to the groundtruth, with sharp boundaries and smooth surfaces being preserved. Note that, the improvement of our method is more significant when handling scenes with heavy occlusions (e.g., the nested structures in scene \textit{Boxes}) and large noise (e.g., the bottom dots in scene \textit{Dots}). That is because, our SFE and DS-AFE can disentangle LF images into spatial and angular subspaces, and thus facilitate our network to learn spatial textures and LF structures.

\begin{table}
\caption{Running time achieved by using the ``Shift-and-concat'' approach and the proposed DS-AFE for cost volume construction. The proposed DS-AFE achieves a very fast inference speed and significantly accelerates our DistgDisp.}\label{tab:efficiencyDisp}
\vspace{-0.2cm}
\renewcommand\arraystretch{1.1}
\centering
\begin{tabular}{|l|c|c|}
\hline
 \multirow{2}*{Stages} & \multirow{2}*{\textit{ \textbf{Shift-and-concat}}} & \multirow{2}*{\textit{ \textbf{DS-AFE (ours)}}} \\
 &&\\
 \hline
  feature extraction                & 2.078 ms & 2.107 ms \\
  \rowcolor{shadow}
   \textbf{cost volume construction}   &  \textbf{2708 ms}  &  \textbf{0.935 ms} \\
  cost aggregation                  & 1.050 ms & 1.030 ms \\
  disparity regression             & 0.083 ms & 0.082 ms \\
  \hline
  Total                                    & 2711 ms & 4.154 ms \\
\hline
\end{tabular}
\vspace{-0.2cm}
\end{table}

 \textbf{3) Efficiency:} The running time of our method is ranked first in Fig.~\ref{fig:Screenshot-Disp}. Our method spends only 4 milliseconds on each scene and is much faster than other methods. That is because, our method uses DS-AFE to directly convolve pixels under specific disparities for cost volume construction, and avoids using the time-consuming repetitive feature shifting operation. To demonstrate the high efficiency of our DS-AFE, we introduce a variant by using the ``shift-and-concat'' approach for cost volume construction, and compare the running time of this variant and our DistgDisp on each stage. When using the ``shift-and-concat'' approach, there are 80 views that need to be shifted by 8 disparity levels, resulting in 640 shifting operations. As shown in Table~\ref{tab:efficiencyDisp}, the ``shift-and-concat'' approach  spends 2708 milliseconds for cost volume construction while our DS-AFE spends only 0.935 milliseconds. Consequently, the inference  is significantly accelerated by our DS-AFE, and the effectiveness and efficiency of our disentangling mechanism in LF disparity estimation is well demonstrated.

\begin{figure}
\centering
\includegraphics[width=8.8cm]{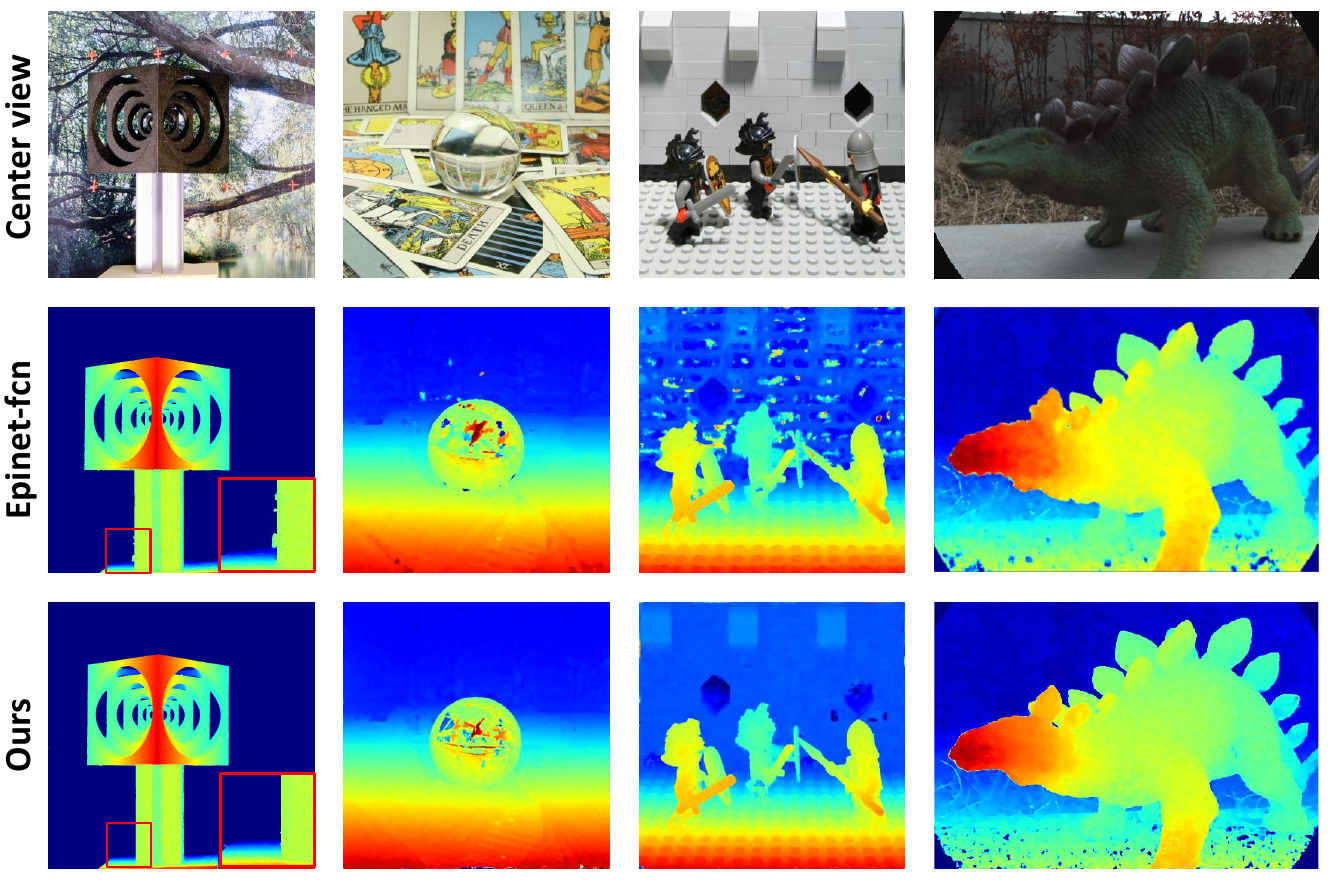}
\caption{Visual results achieved by our method and Epinet-fcn \cite{EPINET} on real-world LFs (from left to right: \textit{Cube} \cite{HCIold}, \textit{Cards} \cite{STFgantry}, \textit{Knights} \cite{STFgantry}, and \textit{Dinosaur} \cite{bok2016geometric}). Our method produces more reasonable disparity maps with sharper edges and fewer artifacts.}
\label{fig:Visual-DispReal}
\end{figure}

\subsubsection{Performance on Real-world LFs and Applications}
We test the performance of our DistgDisp on real-world LFs captured by a moving camera \cite{HCIold,STFgantry} and a Lytro camera \cite{bok2016geometric}. Since the groundtruth disparities are unavailable, we used the model trained on the synthetic LFs for inference, and compared the visual performance of our method to Epinet-fcn \cite{EPINET}. As shown in Fig.~\ref{fig:Visual-DispReal}, the disparity maps produced by our method are more reasonable with sharper edges and fewer artifacts. It demonstrates that our DistgDisp can well generalize to real LFs.

\begin{figure}
\centering
\includegraphics[width=8.8cm]{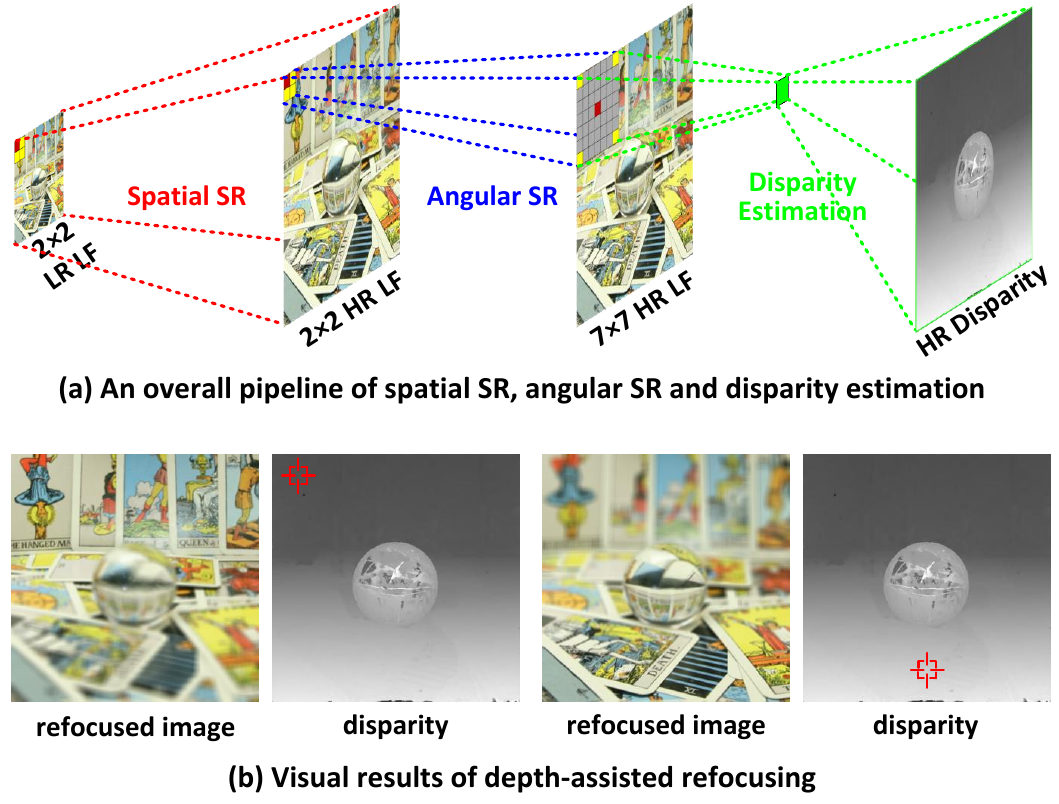}
\caption{Visual results of depth assisted refocusing. (a) By sequentially performing DistgSSR, DistgASR and DistgDisp, an HR dense LF can be reconstructed from an LR sparse LF, and the scene disparities (depths) can be estimated. (b) With the assistance of the estimated disparities, we can refocus LFs to arbitrary regions.}
\label{fig:Visual-Refocus}
\end{figure}

Finally, we apply DistgSSR, DistgASR and DistgDisp sequentially to \textcolor{black}{a 2$\times$2$\times$256$\times$256 LF to achieve 2$\times$ spatial SR, 2$\times$2 to 7$\times$7 angular SR, and disparity estimation.} As shown in Fig.~\ref{fig:Visual-Refocus}(a), with our methods, both spatial and angular resolutions of the input LF are improved and the disparity map can be estimated. Using the estimated disparities, we can further refocus the reconstructed LF onto arbitrary regions. As shown in Fig.~\ref{fig:Visual-Refocus}(b) and \href{https://wyqdatabase.s3.us-west-1.amazonaws.com/DistgLF-demo.mp4}{this video}, the high spatial resolution can provide abundant and faithful details in the focused region while the high angular resolution can provide a natural bokeh effect in the unfocused regions. It demonstrates the practical value of our methods.

\section{Conclusion} \label{sec:Conclusion}
In this paper, we proposed a generic disentangling mechanism for LF image processing. With the proposed spatial, angular and epipolar feature extractors, LFs can be disentangled into different subspace and the inherent structure characteristics can be efficiently learned.
Our disentangling mechanism is compact and can be applied to various LF image processing tasks. Based on our disentangling mechanism, we propose three networks (namely, DistgSSR, DistgASR, and DistgDisp) for three typical LF image processing tasks including spatial SR, angular SR and disparity estimation. Extensive experiments have demonstrated the effectiveness and efficiency of our disentangling mechanism.

\bibliographystyle{IEEEtran}

\bibliography{DistgNet}

\begin{IEEEbiography}[{\includegraphics[width=1in,height=1.25in,clip,keepaspectratio]{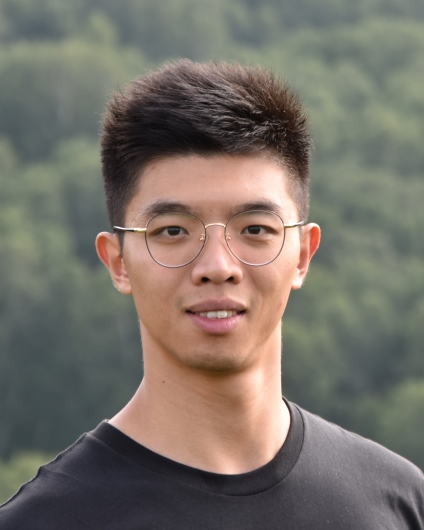}}]{Yingqian Wang} received the B.E. degree in electrical engineering from Shandong University (SDU), Jinan, China, in 2016, and the M.E. degree in information and communication engineering from National University of Defense Technology (NUDT), Changsha, China, in 2018. He is currently pursuing the Ph.D. degree with the College of Electronic Science and Technology, NUDT. His research interests focus on low-level vision, particularly on light field imaging and image super-resolution.
\end{IEEEbiography}

\begin{IEEEbiography}[{\includegraphics[width=1in,height=1.25in,clip,keepaspectratio]{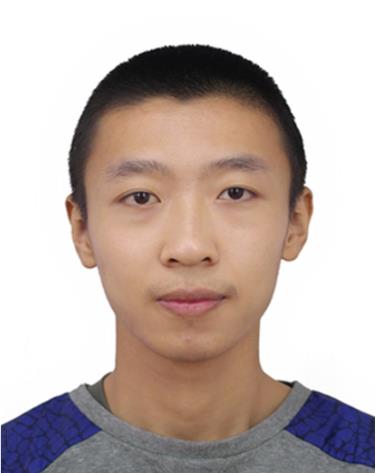}}]{Longguang Wang} received the B.E. degree in electrical engineering from Shandong University (SDU), Jinan, China, in 2015, and the M.E. degree in information and communication engineering from National University of Defense Technology (NUDT), Changsha, China, in 2017. He is currently pursuing the Ph.D. degree with the College of Electronic Science and Technology, NUDT. His research interests include low-level vision and 3D vision.
\end{IEEEbiography}

\begin{IEEEbiography}[{\includegraphics[width=1in,height=1.25in,clip,keepaspectratio]{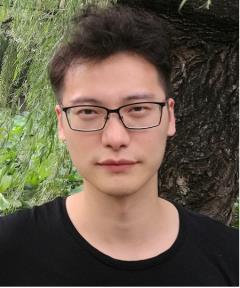}}]{Gaochang Wu} is currently an associate professor at Northeastern  University.  He  received  the  B.E., M.S., and PhD degrees from Northeastern University in  2013, 2015, and 2020,  respectively.  His  research interests  include  image  processing,  light  field  processing and deep learning.
\end{IEEEbiography}

\begin{IEEEbiography}[{\includegraphics[width=1in,height=1.25in,clip,keepaspectratio]{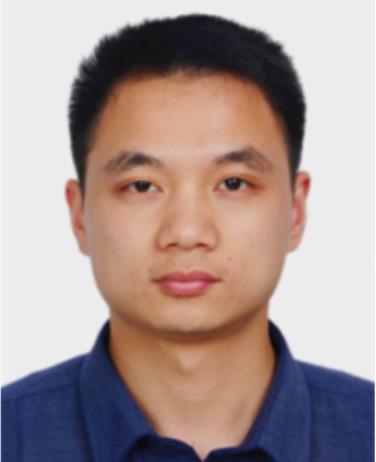}}]{Jungang Yang} received the B.E. and Ph.D. degrees from National University of Defense Technology (NUDT), in 2007 and 2013 respectively. He was a visiting Ph.D. student with the University of Edinburgh, Edinburgh from 2011 to 2012. He is currently an associate professor with the College of Electronic Science, NUDT. His research  interests include computational  imaging, image processing, compressive sensing and sparse representation. Dr. Yang received the New Scholar Award of Chinese Ministry of Education in 2012, the Youth Innovation Award and the Youth Outstanding Talent of NUDT in 2016.
\end{IEEEbiography}

\begin{IEEEbiography}[{\includegraphics[width=1in,height=1.25in,clip,keepaspectratio]{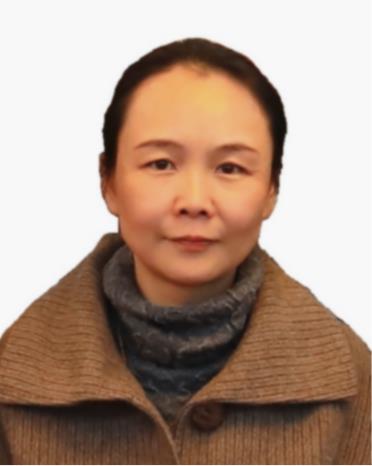}}]{Wei An} received the Ph.D. degree from the National University of Defense Technology (NUDT), Changsha, China, in 1999. She was a Senior Visiting Scholar with the University of Southampton, Southampton, U.K., in 2016. She is currently a Professor with the College of Electronic Science and Technology, NUDT. She has authored or co-authored over 100 journal and conference publications. Her current research interests include signal processing and image processing.
\end{IEEEbiography}

\begin{IEEEbiography}[{\includegraphics[width=1in,height=1.25in,clip,keepaspectratio]{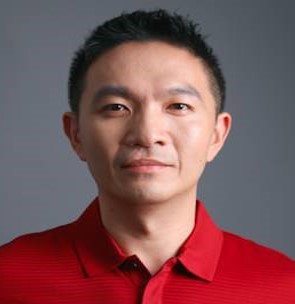}}]{Jingyi Yu} received the BS degree from the California Insitute of Technology, Pasadena, California, in 2000, and the PhD degree from the Massachusetts Institute of Technology, Cambridge, Massachusetts, in 2005. He is currently affiliated with the School of Information Science and Technology, ShanghaiTech University, and the Department of Computer and Information Sciences, University of Delaware. He has authored more than 120 papers at highly refereed conferences and journals, and holds more than 10 international patents on computational imaging. His research interests span a range of topics in computer vision and computer graphics, especially on computational photography and non-conventional optics and camera designs. He is a recipient of the NSF CAREER Award and the AFOSR YIP Award and has served as an area chair of many international conferences including CVPR, ICCV, ICCP, and NeurIPS. He is currently an associate editor of the IEEE TPAMI, the Elsevier CVIU, and Springer IJCV. Dr. Yu is a Fellow of IEEE.
\end{IEEEbiography}

\begin{IEEEbiography}[{\includegraphics[width=1in,height=1.25in,clip,keepaspectratio]{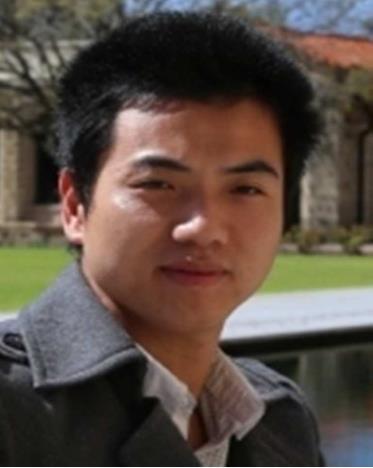}}]{Yulan Guo} is currently an associate professor. He received the B.E. and Ph.D. degrees from National University of Defense Technology (NUDT) in 2008 and 2015, respectively.
He has authored over 100 articles at highly referred journals and conferences. His current research interests focus on 3D vision, particularly on 3D feature learning, 3D modeling, 3D object recognition, and scene understanding. He served as an associate editor for IEEE Transactions on Image Processing, a guest editor for IEEE Transactions on Pattern Analysis and Machine Intelligence, an area chair for CVPR 2021, ICCV 2021, and ACM Multimedia 2021. He organized several tutorials and workshops in prestigious conferences, such as CVPR 2016, CVPR 2019, ICCV 2021, and 3DV 2021. Dr. Guo is a Senior Member of IEEE and ACM.
\end{IEEEbiography}

\end{document}